 \documentclass[final,5p,times,twocolumn,authoryear]{elsarticle}

\usepackage{amssymb}
\usepackage{lipsum}
\usepackage{float} 
\usepackage[hidelinks]{hyperref} 
\usepackage{multirow} 
\usepackage{siunitx}
\usepackage{array}
\usepackage{ragged2e}
\usepackage{booktabs}
\usepackage{multirow}
\usepackage{makecell}
\usepackage{array}
\usepackage{booktabs}
\usepackage{makecell}
\usepackage{multirow}
\usepackage{ragged2e}

\newcolumntype{R}[1]{>{\raggedleft\arraybackslash}p{#1}} 

\usepackage{longtable}
\usepackage{booktabs}
\usepackage[normalem]{ulem}
\usepackage{caption}
\usepackage{ragged2e}
\usepackage{makecell}

\usepackage{array}

\makeatletter
\def\ps@pprintTitle{%
 \let\@oddhead\@empty
 \let\@evenhead\@empty
 \let\@oddfoot\@empty
 \let\@evenfoot\@empty
}
\makeatother

\begin{document}

\begin{frontmatter}



\title{Food safety trends across Europe: insights from the 392-million-entry CompreHensive European Food Safety (CHEFS) database}


\author[first]{N. Kizililsoley\textsuperscript{*}}
\author[first]{F. van Meer, PhD\textsuperscript{*}}
\author[first]{O. Mutlu}
\author[first]{W.F. Hoenderdaal}
\author[first]{R.G. Hobé}
\author[first]{W. Mu, PhD}
\author[first]{A. Gerssen, PhD}
\author[first]{H.J. van der Fels-Klerx, PhD}
\author[second, third]{Á. Jóźwiak, PhD}
\author[fourth]{I. Manikas, PhD}
\author[first]{A. H\"{u}rriyeto\v{g}lu, PhD}
\author[first]{B.H.M. van der Velden, PhD}

\affiliation[first]{
  organization={Wageningen Food Safety Research}, 
  city={Wageningen},
  country={The Netherlands}
}

\affiliation[second]{
  organization={University of Veterinary Medicine Budapest}, 
  city={Budapest},
  country={Hungary}
}

\affiliation[third]{
  organization={Syreon Research Institute}, 
  city={Budapest},
  country={Hungary}
}

\affiliation[fourth]{
  organization={Faculty of Agrobiology, Food, and Natural Resources, Czech University of Life Sciences}, 
  city={Prague},
  country={Czechia}
}

\begin{abstract}
In the European Union, official food safety monitoring data collected by member states are submitted to the European Food Safety Authority (EFSA) and published on Zenodo. This data includes 392 million analytical results derived from over 15.2 million samples covering more than 4,000 different types of food products, offering great opportunities for artificial intelligence to analyze trends, predict hazards, and support early warning systems. However, the current format with data distributed across approximately 1000 files totaling several hundred gigabytes hinders accessibility and analysis. To address this, we introduce the \emph{CompreHensive European Food Safety (CHEFS) database}, which consolidates EFSA monitoring data on pesticide residues, veterinary medicinal product residues, and chemical contaminants into a unified and structured dataset. We describe the creation and structure of the CHEFS database and demonstrate its potential by analyzing trends in European food safety monitoring data from 2000 to 2024. Our analyses explore changes in monitoring activities, the most frequently tested products, which products were most often non-compliant and which contaminants were most often found, and differences across countries. These findings highlight the CHEFS database as both a centralized data source and a strategic tool for guiding food safety policy, research, and regulation.
\end{abstract}




\end{frontmatter}



\begingroup
\renewcommand\thefootnote{}\footnotetext{\textsuperscript{*} These authors contributed equally to this work.}
\endgroup

\section{Introduction}
\label{introduction}

Food safety control includes all activities and actions aimed at ensuring that food is safe for consumption, such as legislation, standards, and food inspection \citep{Sorbo2022}. Food and its ingredients may be contaminated with food safety hazards including chemical and biological hazards. Chemical hazards include (residues of) intentionally applied substances, such as pesticide residues and processing contaminants (e.g., acrylamide), and environmental contaminants (e.g., heavy metals). Microbiological hazards include bacteria, fungi, viruses, parasites, and prions. With a growing global population, ongoing climate change, and an increasingly complex supply chain, ensuring food safety is more important than ever \citep{Fanzo2021,Hu2023}. In response to this growing challenge, we present a comprehensive database of European food safety monitoring data and demonstrate its potential by analyzing trends across Europe from 2000 to 2024.

Maximum limits have been set for a range of food safety hazards for food ingredients and derived products to protect human health. In the European Union (EU), legislation is put in place for official control by member states (Regulation (EU) 2017/625). As a result of this, member states have food safety monitoring activities in place in which they sample and analyze various food safety parameters in food and feed. These parameters include the presence of several chemical and biological hazards. These food safety monitoring activities, whether risk-based to detect non-compliance or random to inform risk assessment, support the overall goal of minimizing risks to public health \citep{Focker2018}.

Results of the food safety monitoring activities collected by member states are sent to the European Food Safety Authority (EFSA). EFSA is an independent agency of the EU (Regulation 178/2002), which compiles and analyzes the data from the food safety monitoring activities to monitor trends, assess risks, and support regulatory decisions. Additionally, EFSA issues calls for national data collection on specific contaminants, such as acrylamide and heavy metals, to refine or start its risk assessments if needed.  The legal basis for the data collection and reporting, including the reporting format, is Article 23 and 33 of Regulation (EC) 178/2002, and the EFSA mandate M-2010-0374.

In order to facilitate reporting of occurrence data to EFSA from several food safety domains and multiple Member States, data need to be reported in a specific, standardized format, called Standard Sample Description (SSD), currently in version 2 (SSD2). The specification of a logical model for SSD2 is composed of: 1) data elements definition and structure; 2) controlled terminologies; and 3) business rules to ensure the validity of the information supplied \citep{EFSA2013}.

The food safety data collected by EFSA is a unique and valuable resource, offering opportunities for research and applications beyond their original purpose. EFSA provides open access to its food safety monitoring data via Zenodo, where the data are published as separate files per member state, hazard category, and year. These files are frequently used in scientific studies \citep{Liu2021,Tarazona2022, Wang2023}. Although these studies show the promise of using EFSA data from Zenodo, they typically use one or a few files. The entire set of files published by EFSA on Zenodo is a vast collection of roughly 392 million analytical results from over 15.1 million samples, covering 4,035 different food products (date: June 10, 2025). This dataset presents a great opportunity for using artificial intelligence to analyze trends, predict future presence of food safety hazards, and develop early warning systems for food safety. However, its current format—split into approximately 1000 files by member state, year, and hazard category and with a combined size of several hundred gigabytes—makes it difficult to use efficiently.

To address this challenge, we introduce the CompreHensive European Food Safety (CHEFS) database. This database integrates all EFSA monitoring data on pesticides, veterinary medicinal product residues (VMPR), and chemical contaminants in food and feed, providing a unified and structured dataset for more accessible and advanced analysis. Other food safety monitoring data submitted by Member States to EFSA—such as data on zoonoses and antimicrobial resistance—are not currently included in this paper due to differences in data structures and metadata. 

In this paper, we describe the structure and creation of the CHEFS database and use it to explore trends in food safety monitoring in Europe between 2000 and 2024. We examine how monitoring activities have evolved over time, identify the food products most frequently tested, and determine which foods and contaminants most often exceed legal safety limits. Finally, we look at how monitoring patterns and outcomes differ between countries. Through these analyses, we demonstrate the potential of the CHEFS database not only as a tool for harmonizing and accessing monitoring data, but also as a resource for identifying priority areas in food safety policy, research, and regulation.

\section{Methods}\label{methods}

\begin{figure*}[t]
    \centering 
	\begin{minipage}{0.8\textwidth}
    \includegraphics[width=\textwidth]{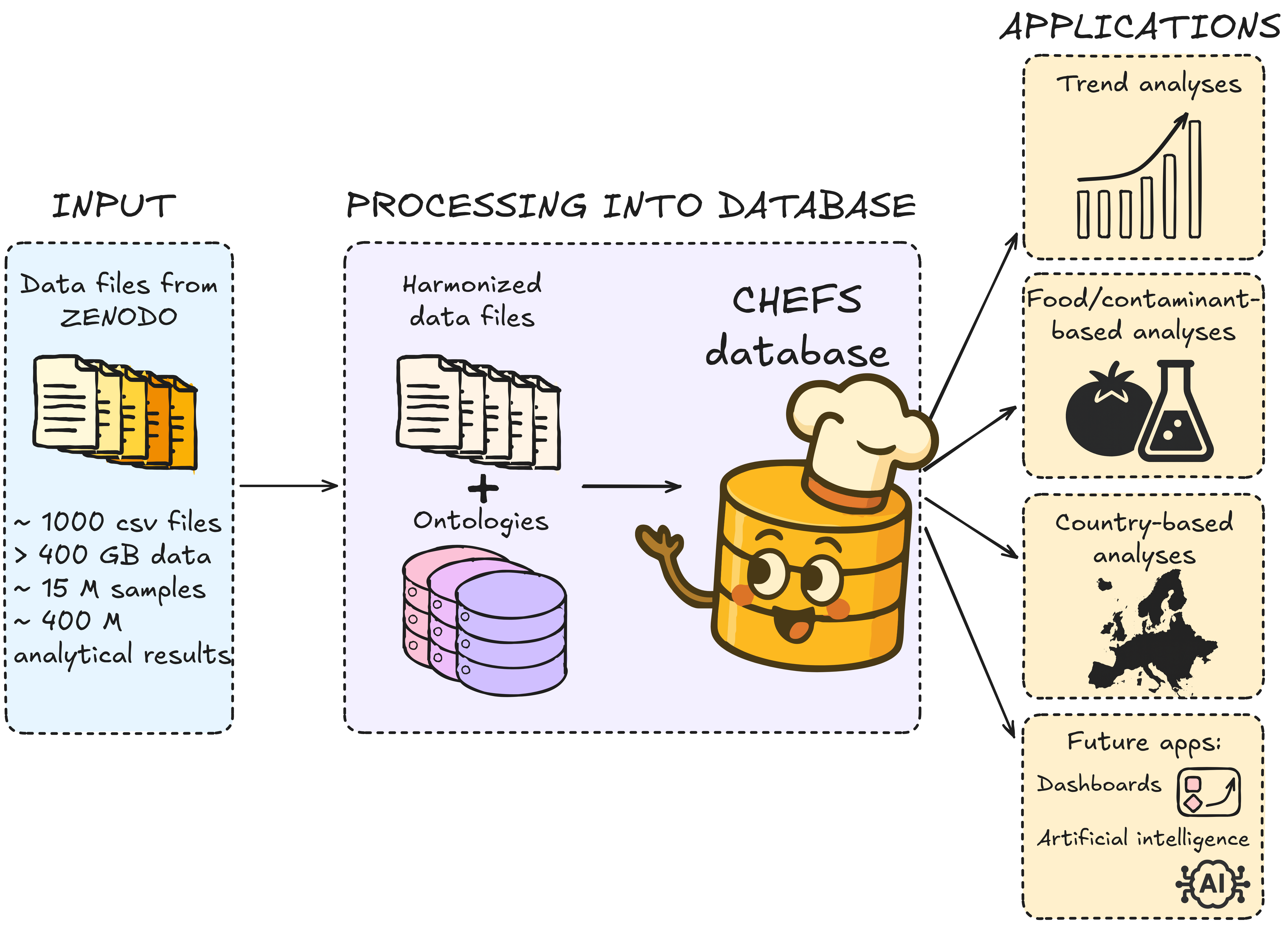}	
	\caption{Schematic overview of the CHEFS database construction and applications presented in this paper.} 
	\label{fig1}
    \end{minipage}
\end{figure*}

In this section we first describe the creation of the CHEFS database and
then describe the statistical analyses that we performed (\autoref{fig1}).

\subsection{Creation of the CHEFS database}\label{creation-of-the-chefs-database}

\subsubsection{Structure of EFSA data files}\label{structure-of-efsa-data-files}

The EFSA data files on Zenodo are structured as a large table in CSV format, where each row represents an individual analytical result, and when multiple analyses were performed in one sample, sample-related information is repeated across multiple rows (i.e., long format). The data follows a one-to-many (1:n) relationship, where a single sample can have multiple analytical results, but each analytical result corresponds to only one sample.

Each sample contains metadata related to the analyzed food item, including product name, country of origin, and sampling date. Additionally, each sample is associated with multiple analyses, capturing information such as the analyzed contaminant (e.g., arsenic, aflatoxin B1, acrylamide), the analytical result, the date of analysis, the limit of quantification (LOQ) of the applied method, and approximately 80 other variables related to the sample and the analysis in the SSD format. Examples are variables that provide information on the size of the sample, its percentage of fat and other attributes, the method or instrument used to analyze the contaminant, whether the analytical result has been corrected for recovery, and so forth.

EFSA categorizes the chemical hazard data into three main groups based on the type of contaminants analyzed: pesticides, chemical contaminants\footnote{Chemical contaminants are defined by EFSA (\emph{Chemical Contaminants in Food and Feed \textbar{} EFSA}, n.d.) as substances that are not intentionally added to food or feed but are present as a result of environmental contamination (e.g. heavy metals like lead, cadmium, mercury), food processing (e.g. acrylamide, polycyclic aromatic hydrocarbons), packaging or contamination during production, transport, or storage.} (e.g., heavy metals, mycotoxins,
dioxins), and veterinary medicinal product residues (VMPR).

\subsubsection{Data processing}\label{data-processing}

Data processing focused on harmonizing data formats, identifying unique identifiers for samples and analytical results, handling missing values, and removing duplicates. There were separate datafiles per contaminant group, country, and year. The development of the CHEFS database followed a structured, step-by-step process. It began with the automatic downloading and extraction of EFSA's raw data files (as described in 2.1.1). The next step involved analyzing and harmonizing the varying data formats---not only between the main contaminant groups (pesticides, chemical contaminants, and veterinary medicinal product residues), but also within them. For example, even within the VMPR category, several different formats were used that differed from each other and from those of the other contaminant types. Such differences include different column names for the same variables, not all variables present in all datasets, or variables with only missing values in one file but (partially) filled with data in another file.

With the basic structure in place, we identified variables related to samples and variables related to analytical results. To better understand the quality and usability of the data, we assessed data sparsity by investigating the missing value rates per variable. Subsequently, we linked EFSA's reference catalogues -- i.e., SSD1 or SSD2, depending on the year of reporting -- to the data, enriching it with standardized metadata. These reference catalogues include contaminant ontologies, product classifications, and country codes. After these steps, we designed the database scheme and created a relational database.

To enable efficient and repeatable data integration, scripts for automated processing of the raw EFSA data files and importing them into the CHEFS database were developed. Due to the variations between the raw data files, the processing had to be done with care. For example, raw data files could have different variables, have the same variables but with different names, or, as in the case of product information, use different ontologies based on the year. Finally, the entire system was rigorously validated: data exported from the CHEFS database were compared to the original Zenodo source files to ensure that no information had been altered or lost during processing.

\subsubsection{\texorpdfstring{CHEFS database storage design }{CHEFS database storage design }}\label{chefs-database-storage-design}

The CHEFS database is a PostgreSQL relational database, structured according to the sample/analytical result model (\autoref{figA1}). To address the high number of variables and the sparsity of data (\autoref{figA2}), the database design incorporated ``core'' and ``rest'' tables. Variables including essential information to the sample and/or analytical result were placed in the ``core'' table, all other sample variables were stored in the ``rest'' table.

Several predefined data selections were created to simplify data retrieval by combining information from related tables. This was done to simplify running queries on the CHEFS database. Furthermore, we provided a Python script to automate data extraction and merging, which streamlines dataset construction for analysis.

\subsection{Statistical analysis}\label{statistical-analysis}

Descriptive statistical analyses were performed for three main types of analyses: trends over years; contaminant and food statistics; and country-based statistics.

We defined an analytical result to be above limits or ''non-compliant'' when the evaluation code (variable name evalcode\_id) was either ``Greater Than Maximum Permissible Quantities'', ``Non-Compliant'', ``Detected'', or ``Unsatisfactory'' (\autoref{tab:eval-results}). 
\begin{table}[ht]
\footnotesize
\captionsetup{width=\columnwidth}
\caption{Number and percentage of analytical results in the CHEFS database. Analytical results with the evaluation codes “Greater Than Maximum Permissible Quantities”, “Non-Compliant”, “Detected”, and “Unsatisfactory” were defined as above limits or ‘non-compliant’.}
\label{tab:eval-results}
\begin{center}
\begin{tabular}{p{2cm} p{3.5cm} r r}
\toprule
\textbf{Evaluation of analytical result value} & 
\textbf{Analytical results} & 
\textbf{\% of total} \\
\midrule
Less than or equal to max permissible quantities & 270,592,813 & 68.98\% \\
Result not evaluated & 94,197,203 & 24.01\% \\
Not detected & 25,580,634 & 6.52\% \\
Compliant & 1,445,071 & 0.37\% \\
Compliant due to measurement uncertainty & 351,142 & 0.09\% \\
\textit{Greater than max permissible quantities} & \textit{59,549} & \textit{0.02\%} \\
\textit{Detected} & \textit{35,274} & \textit{0.01\%} \\
Acceptable & 3,230 & 0.00\% \\
Satisfactory & 2,564 & 0.00\% \\
\textit{Non-compliant} & \textit{2,401} & \textit{0.00\%} \\
\textit{Unsatisfactory} & \textit{30} & \textit{0.00\%} \\
\midrule
\textbf{TOTAL} & \textbf{392,269,911} & \textbf{100.00\%} \\
\bottomrule
\end{tabular}
\end{center}
\end{table}

All statistical analyses and visualizations were done using Python. Datasets used in the analyses and visualizations were extracted from the CHEFS database using psycopg2, a PostgreSQL database adapter in Python. Scripts for querying the database and creating the visualizations are available at \url{https://github.com/WFSRDataScience/CHEFS}.

\subsubsection{Trends over years}\label{trends-over-years}

We examined trends over time in food safety monitoring between 2000-2024 using data from the CHEFS database.\footnote{EFSA database has five samples from years earlier than 2000. For analyses that did not incorporate year information, we included all available data.} The data included results from different hazard categories---chemical contaminants, pesticide residues, and VMPR, across various food product types (whether it be food product or a matrix involved in food processing) and sampling countries. We calculated the number of analytical results and the number of analytical results above their respective legal limits, per year, per hazard category, and visualized them with line charts.

\subsubsection{Contaminant- and food-based statistics}\label{contaminant--and-food-based-statistics}

\subparagraph{Contaminant-based statistics}\label{contaminant-based-statistics}

A total of 4,170 contaminant IDs were included in the CHEFS database, spanning chemical contaminants, pesticide residues, and VMPR. We examined the trends in most frequently tested hazards by aggregating all analytical results associated with each unique contaminant ID across countries, years, and product types per contaminant type. The 10 most frequently analyzed contaminants were visualized using bar charts, stratified by hazard category (chemical contaminants, pesticide residues, and VMPR).

To explore hazard-product associations, we identified the 10 most frequently measured hazards for each contaminant type, based on total analytical result counts. For each hazard, we showed the top 10 products in which they were most frequently measured, and also showed the percentage of non-compliance.

We grouped the contaminant identifiers (IDs) using the top-level categories assigned in the standard knowledge system (i.e., an ontology). An ontology is a structured framework that defines concepts and their relationships within a specific domain, organized hierarchically from general to specific to enable shared understanding and interoperability \citep{Gruber1993}. The contaminant names come from the PARAM (Parameter) controlled terminology developed by EFSA. For instance, the full name for the hazard ``cadaverine'' is provided by EFSA database as ``toxins::biogenic amines::cadaverine''. We parsed full names by using ``::'' as a separator, and grouped each hazard by the first term in its full name, categorizing cadaverine under the head group ``toxins''. After grouping contaminant IDs by their first level ontologies \citep{vanderFels-Klerx2024}, 10 main contaminant groups were visualized in a bubble chart.

\subparagraph{Food-based statistics}\label{food-based-statistics}

The CHEFS database contains a total of 4,035 product IDs measured across chemical contaminants, pesticide residues, and VMPR. To analyze product-hazard associations, a table was constructed similar to the hazard-product summary described above. This analysis highlights the most commonly monitored product-contaminant relationships in the dataset.

With the intention of dealing with same food products being scattered to vast amount of different classification of products, and to get a more compact understanding of product-based statistics, individual product IDs were again grouped into broader food categories\footnote{For example, pesticide residue measurement records for hen eggs can be found under either of the names "matrix::birds eggs", "matrix::eggs (chicken)", "mtx::boiled eggs", "mtx::egg yolk", "mtx::hen egg mixed whole", "mtx::hen egg mixed whole, dried", "mtx::hen egg yolk", "mtx::hen egg yolk, dried", "mtx::hen eggs", "mtx::liquid egg products", "mtx::whole eggs". Therefore we decided to do series of logical grouping with parsing full names of products and assigning them to logical groups via keyword search. This way, ``mtx::whole eggs'' would be assigned to a higher group of ``Eggs and egg products'', from its full name ``mtx::all lists::food::eggs and egg products::whole eggs''.}. 
This scheme also helps for dealing with nonspecific or miscellaneous names in the database. Although hierarchical classification systems were available for each product ID from EFSA, differences exist over the years. This is, for example, because of the switch from the FoodEx to the FoodEx2 matrix catalog and subsequent updates to the catalog \citep{EFSA2025}. Therefore, we build a custom grouping strategy on top of these EFSA classification systems. To do this, we analyzed product names at their full hierarchical labels, and a series of logical rules were used to assign each product to a standardized category based on their position within the hierarchy and their type (chemical contaminants, pesticide residues, or VMPR). Categories such as "Fruits", "Cereals", "Animal meat and tissues", and "Milk and milk products" were defined based on keywords and patterns within the analyzed product names. These groupings enabled aggregation and visualization of contamination data at a meaningful categorical level.

Bar plots were generated separately for the three hazard categories (chemical contaminants, pesticide residues, and VMPR), showing the ten most frequently measured categories for each group, with both numbers of total analytical results and analytical results above legal limits displayed. The full dictionary for this grouping is given in \autoref{tableA1}.

\subsubsection{Country-based statistics}\label{country-based-statistics}

Within the EFSA database, every sample analyzed is associated with details regarding its ``origin country'' (where the product was produced) and ``sampling country'' (where the sample was collected for testing). To investigate the total analytical results and non-compliant results per sampling country and contaminant type, we displayed the top 15 countries based on number of analytical results per contaminant type (chemical contaminants, pesticide residues, and VMPR) with bar charts.

Having the information of sample origins, we also examined trade flow using chord diagrams, which represent product movements from origin to destination countries. Each origin country-sampling country pair was treated as one linkage. We filtered out self-referencing country pairs and the pairs with few samples (results for \textgreater{} 100 and \textgreater{} 5000 samples are shown). The top 20 links between origin countries and sampling country were shown.

\section{Results}
\subsection{Overview of the data}
In total, 35 European Union Member and Associated States\footnote{See the list of EU Member States, Members of the Advisory Forum, and Observers for EFSA on \url{https://www.efsa.europa.eu/en/partnersnetworks/eumembers}.} reported 392,269,911 analytical results in 15,176,473 samples, collected between 2000-2024. Of these 392 million analytical results, 97,254 were ‘non-compliant’ (0.025\%) (\autoref{tab:eval-results}). Analyzed samples originated from 235 different countries. 

Among the 15 million samples included in the CHEFS data, around 9 million samples were analyzed only for one hazard, 2,413,837 were analyzed for 13 hazards, and between 10-10,000 of them had an average of 286.8 analytical results per sample (\autoref{figA3}). 

Of all analytical results, 1.11\% were for chemical contaminants (in 5.96\% of samples), 78.16\% for pesticide residues (in 57.92\% of samples), and 20.74\% for VMPR (in 36.12\% of samples). The average number of analytical results per sample is 4.8 for chemical contaminants, 34.9 for pesticide residues, and 14.8 for VMPR (\autoref{tableA2}). It should be noted that some samples are analyzed for multiple types of hazards.

Sampling strategies included objective (random) sampling in 53.1\% of the samples, selective (risk-based) sampling in 23.4\%, suspect sampling in 5.3\%, and 18.0\% was labeled as “Other”. Convenient sampling, and “Not Specified” covered 0.1\% of the total number of samples (\autoref{tableA3}).
The CHEFS database contained 3,180 contaminants and 4,035 unique product IDs.

\subsection{Trends over years}
\begin{figure*}[htbp]
    \centering 
    \begin{minipage}{0.8\textwidth}
	\includegraphics[width=\textwidth]{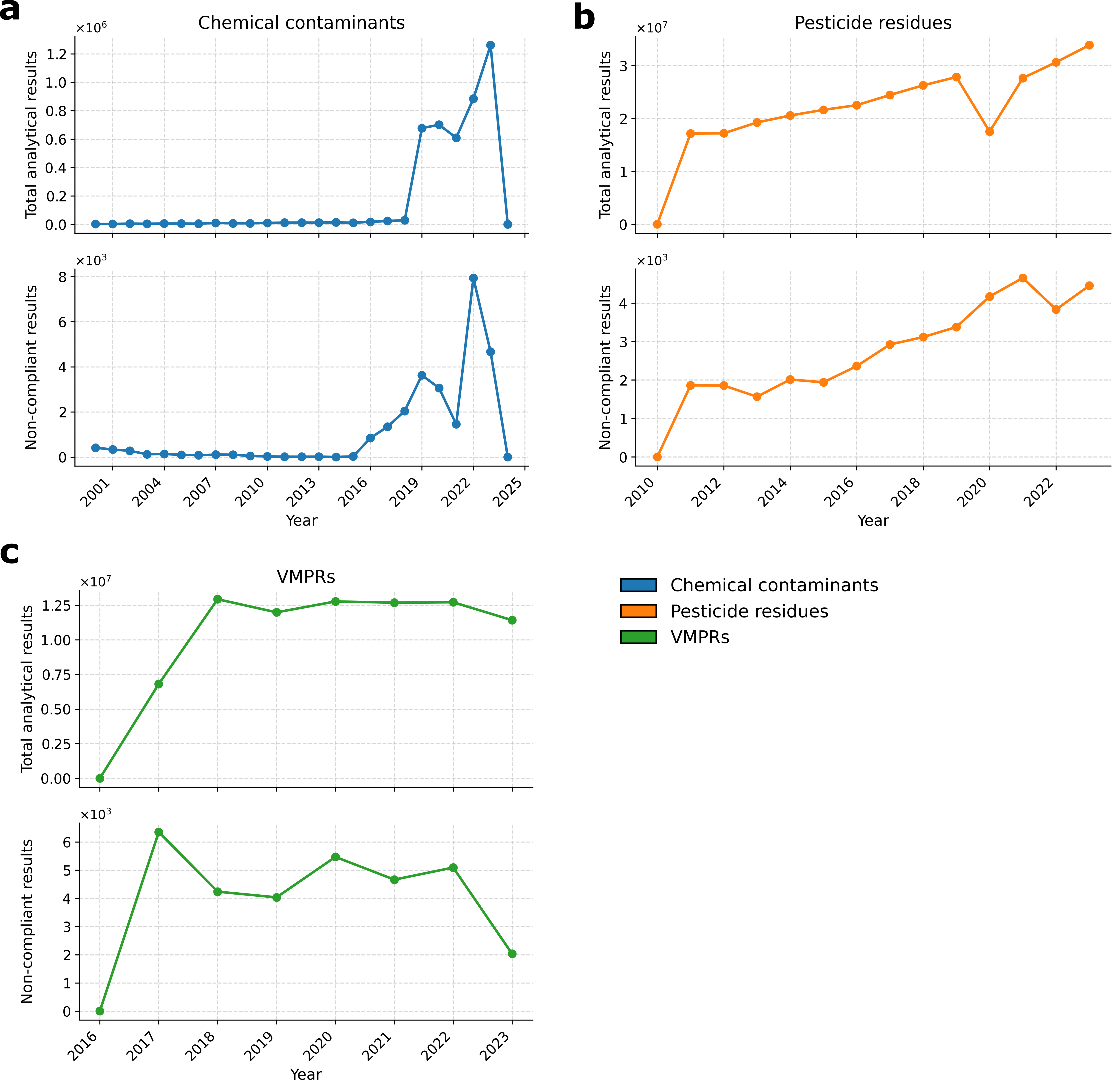}	
	\caption{Number of analytical results and number of analytical results above legal limits of (a) chemical contaminants,  (b) pesticide residues, and (c) veterinary medicinal product residues (VMPR). All groups show an increase over time in analytical result, and pesticide residues shows an increase over time in analytical results above limit. It should be noted that each graph uses different scales, with axes annotated using scientific notation (powers of 10), as indicated next to each axis.} 
	\label{fig2}
    \end{minipage}   
\end{figure*}

The number of analytical results performed for chemical contaminants, pesticide residues, and VMPR, showed a general increase over time (\autoref{fig2}). We included the monitoring data of after 2000, even though available data by EFSA starts from 1970. While analytical results of chemical contaminants have been recorded since 2000 in the EFSA database, pesticide residue monitoring data was only available after 2010, and VMPR was only available after 2016. 

The number of analytical results of chemical contaminant group increased after 2016 (\autoref{fig2}). The percentage of analytical results that surpassed the legal limits also increased, from 0.2\% in 2015 to 4.6\% in 2016, 5.6\% in 2017, and 6.9\% in 2018. The distribution of these numbers over different sampling strategies between 2015-2019 is given in \autoref{tableA4}. In 2019, the total number of analytical results in the chemical contaminant group increased from 29,415 to 677,137, while the percentage of analytical results above legal limits dropped to 0.5\%. This may be partially explained by differences in sampling strategies over the years. For example, of all analytical results reported for chemical contaminants in the year 2018, 41\% were objective sampling, 41\% selective sampling, 17\% convenient sampling, and 1\% suspect sampling. In 2019, however, 55\% of the analytical results were based on selective sampling, 32\% objective sampling, 9\% suspect sampling and 2\% convenient sampling. The percentages of non-compliance for this year were 0.24\% for selective sampling, 0.19\% for objective sampling, 0.55\% suspect sampling, 12.31\% for convenient sampling and 0.48\% for “other”.

The number of analytical results in pesticide residues consistently increased after 2011, although there was a sharp drop in the total amount of analytical results for 2020 (\autoref{fig2}). This is because there were very few records available for Germany, Lithuania, and Ireland for 2020. This could have been due to the effects of COVID-19 lockdowns, as inspectors were in some cases/countries not allowed to visit farms or companies. The number of analytical results above legal limits increased over time, also between 2019-2021. 
The number of analytical results in VMPR remained stable after 2018 (\autoref{fig2}). The percentage of non-compliant cases was the highest in 2017, with 6,356 out of 6,813,496 (0.093\%) analytical results above legal limits.

\subsection{Contaminant- and food-based statistics}
The CHEFS database provides a detailed insight on the distribution of different types of food and contaminant monitoring.

\subsubsection{Contaminant-based statistics}
While analyzing contaminant-based statistics, it is important to note that of the 4,529 contaminant IDs in the dataset, only 3,180 were unique. This is because some contaminant IDs were associated with multiple types of analytical results and were therefore not exclusive to a single result type: 2.3\% (n = 73) of the unique contaminant IDs included in all three hazard categories (i.e. chemical contaminants, pesticide residues and VMPR) and 37.8\% of them were used by two of the three categories. Pesticide residues and VMPR had most overlap (\autoref{fig4}).

\begin{figure*}[htbp]
    \centering 
	\begin{minipage}{0.8\textwidth}
    \includegraphics[width=\textwidth]{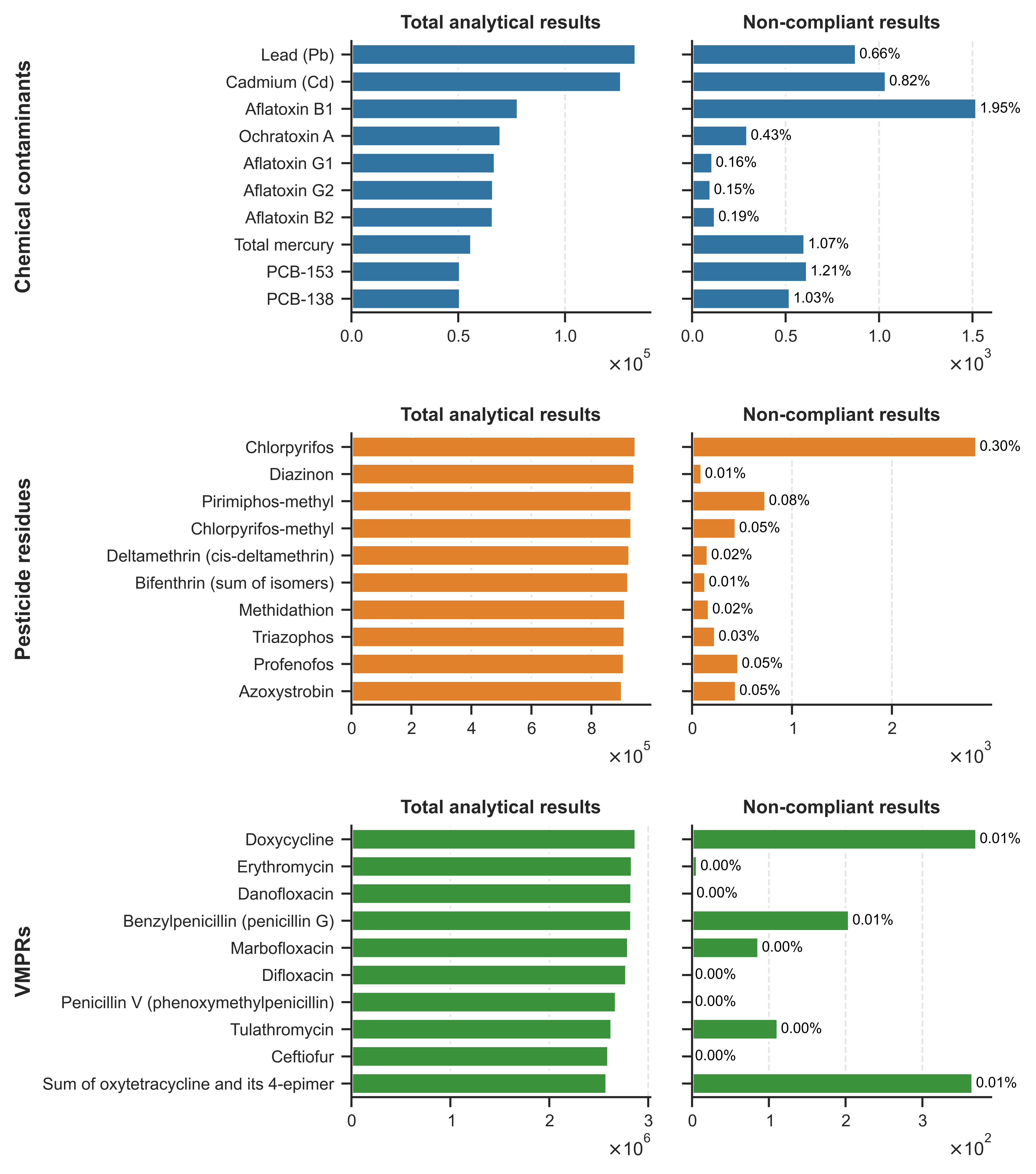}	
	\caption{The top 10 most analyzed contaminants for each hazard group for a) chemical contaminants, b) pesticide residues, and c) veterinary medicinal product residues (VMPR), aggregated over all countries and years. The percentages surpassing the legal limits are shown in the right bar plots. It should be noted that each graph uses different scales, with axes annotated using scientific notation (powers of 10), as indicated next to each axis.} 
	\label{fig3}
    \end{minipage}
\end{figure*}

\begin{figure*}[htbp]
\centering 
    \begin{minipage}{0.8\textwidth}
    \includegraphics[width=\textwidth]{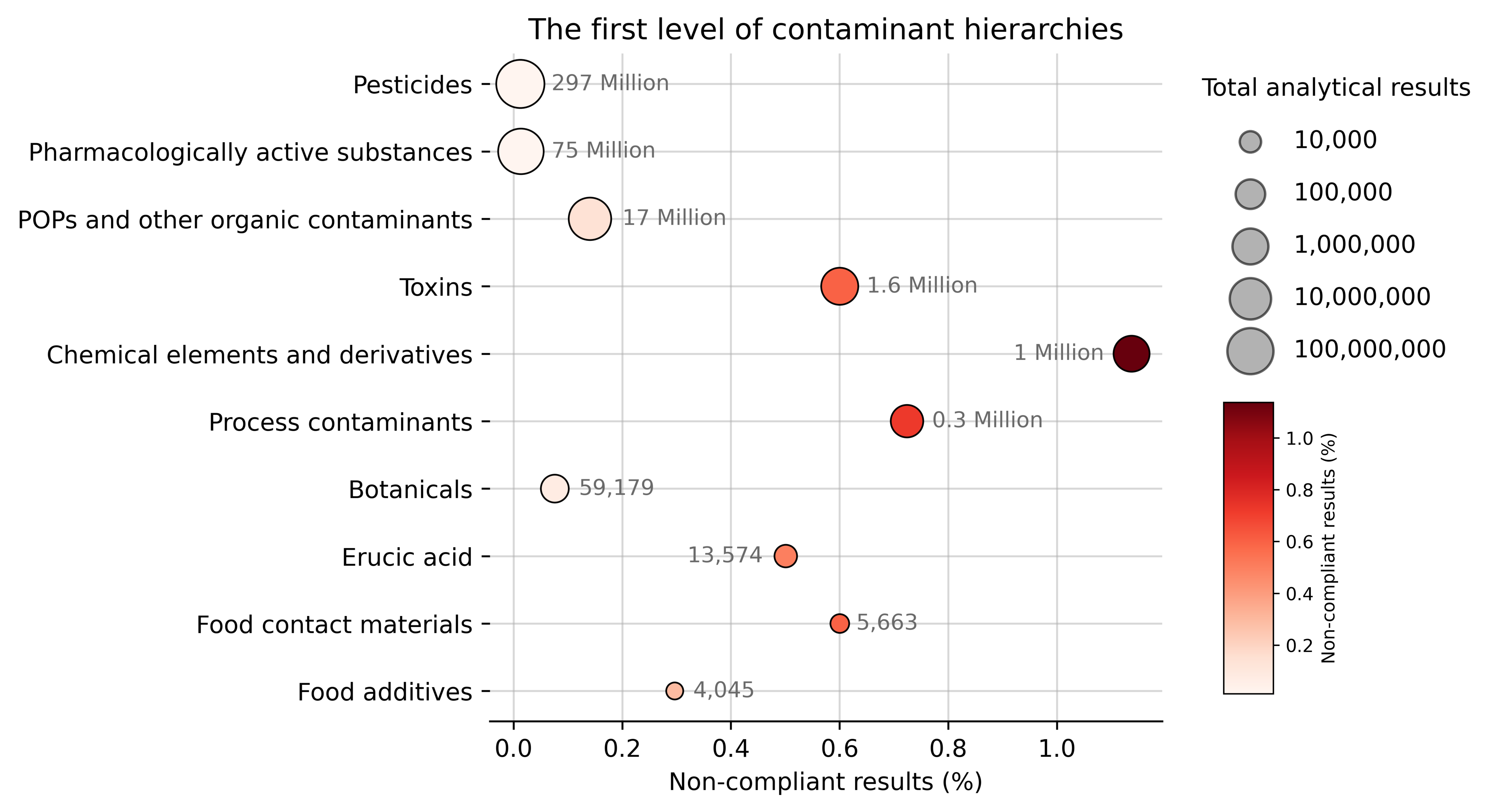}	
	\caption{Main contaminant groups showing the number of analytical results and percentage of analytical results above legal limits. The size of each bubble is scaled logarithmically according to the total number of analytical results collected in that category (ranging from ~4,000 to almost 300 million). The bubble color reflects the percentage of analytical results above legal limits.} 
	\label{fig4}
    \end{minipage}
\end{figure*}

The top 10 most analyzed contaminants per category are shown in \autoref{fig3}. Within chemical contaminants, lead (Pb) was the most tested contaminant with 133,264 analytical results (3.1\% of the 4,344,679 total analytical resutls for 897 unique contaminants in chemical contaminants group), followed by cadmium (Cd) (2.9\%), and aflatoxin B1 (1.8\%) (\autoref{table2}). Pig fresh meat, pig liver, and “cow, ox or bull fresh meat” were the top 3 most analyzed products for lead (Pb) and cadmium (Cd). For lead (Pb), non-compliant analytcial results were observed at rates of 0.58\% (out of 4,276), 0.24\% (out of 3,355), and 0\% (out of 2,612), respectively. For cadmium, these rates were 0.59\% (out of 4,088), 0.42\% (out of 3,344), and 0.08\% (out of 2,595). Aflatoxin B1 was most frequently analyzed in peanuts, pistachios, and dried figs, with non-compliance rates of 4.01\% out of 17,292, 4.14\% out of 6,231, and 2.70\% out of 4,736 analytical results, respectively.

\begin{table*}[htbp]
\centering
\caption{Top three products for top three hazards in chemical contaminants, selected by highest number of analytical results.}

\footnotesize
\begin{tabular}{p{2cm} p{3.5cm} r r}
\toprule
\textbf{Hazard name}                   & \textbf{Product name}               & \textbf{Total analytical results} & \textbf{Non-compliance ratio} \\
\midrule
\multirow{3}{*}{Lead (Pb)}& Pig fresh meat             & 4,276                    & 0.58\%               \\
                              & Pig liver                  & 3,355                    & 0.24\%               \\
                              & Cow, ox or bull fresh meat & 2,612                    & 0.00\%               \\
                              \midrule
\multirow{3}{*}{Cadmium (Cd)}& Pig fresh meat             & 4,088                    & 0.59\%               \\
                              & Pig liver                  & 3,344                    & 0.42\%               \\
                              & Cow, ox or bull fresh meat & 2,595                    & 0.08\%               \\
                              \midrule
\multirow{3}{*}{Aflatoxin B1} & Peanuts                    & 17,292                   & 4.01\%               \\
                              & Pistachios                 & 6,231                    & 4.14\%               \\
                              & Dried figs                 & 4,736                    & 2.70\%   \\          
\bottomrule

\end{tabular}
\label{table2}
\end{table*}

For 1,612 pesticide residues, the top tree contaminants were chlorpyrifos with 949,017 analytical results (0.31\%), diazinon 944,958 (0.31\%), and pirimiphos-methyl 935,180 (0.30\%) (\autoref{table3}). When we ordered products analyzed for chlorpyrifos by the total number of analytical results, “sweet peppers/bell peppers”, “sweet peppers”, and “apples” appear on top 3, with non-compliance rates of 0.48\% (out of 34,548 analytical results), 0.21\% (out of 33,610), and 0.27\% (out of 23,723). These products were also the top 3 most analyzed items for diazinon and pirimiphos-methyl. Readers should be aware that this repetition of sweet peppers in the list may have originated from the fact that there are two different classification systems under EFSA’s repository for pesticide residue analytical results: product names from FoodEx2 as “mtx” and EFSA’s harmonized commodity codes as “matrix”.

\begin{table*}[htbp]
\centering
\caption{Top three products for top three hazards in pesticide residues, selected by highest number of analytical results.}

\footnotesize
\begin{tabular}{p{2cm} p{3.5cm} r r}
\toprule
\textbf{Hazard name}          & \textbf{Product name}         & \textbf{Total analytical results} & \textbf{Non-compliance ratio} \\
\midrule
\multirow[t]{3}{*}{Chlorpyrifos}      
  & Sweet peppers/bell peppers & 34,548 & 0.48\% \\
  & Sweet peppers              & 33,610 & 0.21\% \\
  & Apples                     & 23,723 & 0.27\% \\
\midrule
\multirow[t]{3}{*}{Diazinon}           
  & Sweet peppers              & 35,432 & 0.00\% \\
  & Sweet peppers/bell peppers & 34,162 & 0.01\% \\
  & Apples                     & 23,337 & 0.00\% \\
\midrule
\multirow[t]{3}{*}{Pirimiphos-methyl}  
  & Sweet peppers              & 35,306 & 0.30\% \\
  & Sweet peppers/bell peppers & 34,300 & 0.92\% \\
  & Apples                     & 23,207 & 0.00\% \\
\bottomrule
\end{tabular}
\label{table3}
\end{table*}

The contaminant within the VMPR category with the highest number of analytical results was doxycycline with 2,877,150 analytical results (3.54\%), followed by erythromycin with 2,838,678 (3,49\%), and danofloxacin 2,834,309 (3.48\%) (\autoref{table4}). Out of 2,877,150 analytical results available for doxycycline, 1,509,230 was done for pig kidney, with ~0\% rate of non-compliance, followed by pig fresh meat with 0.02\% and “cow, ox or bull fresh meat” with ~0\%. Pig kidney, pig fresh meat, and “cow, ox or bull fresh meat” were also the top 3 most analyzed products for erythromycin and danofloxacin, where all non-compliance rates were close to 0\%. 

The full list of top 10 products per top 10 contaminants in \autoref{fig3}, ordered by the total number of analytical results are given in \autoref{tableA4}. 

\begin{table*}[htbp]
\centering
\caption{
Top three products for top three hazards in veterinary medicinal product residues (VMPR), selected by highest number of analytical results.}

\footnotesize
\begin{tabular}{p{2cm} p{3.5cm} r r}
\toprule
\textbf{Hazard name}          & \textbf{Product name}         & \textbf{Total analytical results} & \textbf{Non-compliance ratio} \\
\midrule
\multirow[t]{3}{*}{Doxycycline}      
  & Pig kidney                 & 1,509,230 & 0.00\% \\
  & Pig fresh meat            &   789,576 & 0.02\% \\
  & Cow, ox or bull fresh meat&   102,206 & 0.00\% \\
\midrule
\multirow[t]{3}{*}{Erythromycin}     
  & Pig kidney                 & 1,507,251 & 0.00\% \\
  & Pig fresh meat            &   785,711 & 0.00\% \\
  & Cow, ox or bull fresh meat&    99,242 & 0.00\% \\
\midrule
\multirow[t]{3}{*}{Danofloxacin}     
  & Pig kidney                 & 1,495,131 & 0.00\% \\
  & Pig fresh meat            &   785,788 & 0.00\% \\
  & Cow, ox or bull fresh meat&    99,065 & 0.00\% \\
\bottomrule
\end{tabular}
\label{table4}
\end{table*}

To have a broader look, the the contaminant ontology groups were also visualized. The contaminant groups “chemical elements and derivatives” and “process contaminants” showed the highest percentage of non-compliant analytical results (\autoref{fig4}). To further investigate this, their subgroups were plotted (\autoref{figA5}). In the subgroup “persistent organic pollutants (POPs) and other organic contaminants” the highest non-compliant analytical result rate was observed for dioxins and PCBs (\autoref{figA5}). In the subgroup “toxins” the highest non-compliant rate was for “phytotoxins” (\autoref{figA5}). In the subgroup “heavy metal elements”, copper had the highest non-compliant analytical result percentages (\autoref{figA5}).

\subsubsection{Food product statistics}
There were 4,035 unique product IDs in the CHEFS database (6,587 in total), analyzed for the wide range of chemical contaminants, pesticide residue, and VMPR analytical results.
Peanuts, hen eggs, and pig fat tissue were the top three products that were analyzed the most for hazards under the chemical contaminants category. Of the 168,152 tests done on peanuts, aflatoxins B1, B2, G1, and G2 were the most commonly tested. The levels of these toxins were found to be 4.01\%, 0.04\%, 0.01\%, and 0.01\%, respectively, of peanut analytical results that were non-compliant. Hen eggs and pig fat tissue were frequently measured for several kinds of PCBs (\autoref{table5}). The most common hazards measured for the hen eggs were PCB-101, PCB-138 and PCB-153. In the pig fat tissue, the most common hazards were PCB-52, PCB-180 and PCB-28.

For pesticide residues, sweet peppers (mtx) have been analysed most frequently, with a total of 9,174,299 analyses recorded. This is followed by sweet peppers/bell peppers (matrix) which have been analysed 8,328,258 times, and apples which have been analysed 7,333,873 times. While sweet peppers (mtx) were most extensively measured for boscalid, cyproconazole, and oxadixyl; sweet peppers/bell peppers were measured for procymidone, oxamyl, and tetradifon (\autoref{table6}). Apples were measured most frequently for chlorpyrifos, penconazole (sum of constituent isomers), and chlorpyrifos-methyl. 

For VMPR, the product that were tested the most was pig kidney, and most frequently for doxycycline, erythromycin, and benzylpenicillin (penicillin G). This was followed by pig fresh meat and non-food animal-related matrices (\autoref{table7}). 

\begin{table*}[htbp]
\centering
\caption{Top three hazards for top three products in chemical contaminants, selected by highest number of analytical results.}
\footnotesize
\begin{tabular}{p{2 cm} R{2cm} p{3.5 cm} R{2cm} R{2cm}}
\toprule
\textbf{Product name} & \textbf{Total analytical results per product} &
\textbf{Hazard name} & \textbf{Total analytical results per hazard} &
\textbf{Non-compliance ratio} \\
\midrule
\multirow[t]{3}{=}{Peanuts}
  & 168,152 & Aflatoxin B1 & 17,292 & 4.01\% \\
  &         & Aflatoxin B2 & 16,713 & 0.04\% \\
  &         & Aflatoxin G1 & 16,712 & 0.01\% \\
\midrule
\multirow[t]{3}{=}{Hen eggs}
  & 157,649 & PCB-101 & 4,625 & 0.54\% \\
  &         & PCB-138 & 4,623 & 0.52\% \\
  &         & PCB-153 & 4,623 & 0.52\% \\
\midrule
\multirow[t]{3}{=}{Pig fat tissue}
  & 124,652 & PCB-52  & 5,093 & 0.12\% \\
  &         & PCB-180 & 5,093 & 0.16\% \\
  &         & PCB-28  & 5,093 & 0.10\% \\
\bottomrule
\end{tabular}
\label{table5}
\end{table*}

\begin{table*}[htbp]
\centering
\caption{Top three hazards for top three products in pesticide residues, selected by highest number of analytical results.}
\footnotesize
\begin{tabular}{p{2 cm} R{2cm} p{3.5 cm} R{2cm} R{2cm}}
\toprule
\textbf{Product name} & \textbf{Total analytical results per product} &
\textbf{Hazard name} & \textbf{Total analytical results per hazard} &
\textbf{Non-compliance ratio} \\
\midrule
\multirow[t]{3}{=}{Sweet peppers (mtx)}
  & 9,174,299 & Boscalid      & 40,284 & 0.00\% \\
  &           & Cyproconazole & 40,182 & 0.01\% \\
  &           & Oxadixyl      & 40,182 & 0.00\% \\
\midrule
\multirow[t]{3}{=}{Sweet peppers/bell peppers (matrix)}
  & 8,328,258 & Procymidone & 44,251 & 0.08\% \\
  &           & Oxamyl      & 44,029 & 0.07\% \\
  &           & Tetradifon  & 43,766 & 0.08\% \\
\midrule
\multirow[t]{3}{=}{Apples}
  & 7,333,873 & Chlorpyrifos                              & 23,723 & 0.27\% \\
  &           & Penconazole (sum of constituent isomers) & 23,544 & 0.00\% \\
  &           & Chlorpyrifos-methyl                      & 23,520 & 0.00\% \\
\bottomrule
\end{tabular}
\label{table6}
\end{table*}

\begin{table*}[htbp]
\centering
\caption{
Top three hazards for top three products in veterinary medicinal product residues (VMPR), selected by highest number of analytical results.}

\footnotesize
\begin{tabular}{p{2 cm} R{2cm} p{3.5 cm} R{2cm} R{2cm}}
\toprule
\textbf{Product name} & {\textbf{Total analytical results per product}} &
\textbf{Hazard name} & {\textbf{Total analytical results per hazard}} &
\textbf{Non-compliance ratio} \\
\midrule
\multirow[t]{3}{=}{Pig kidney}
  & 22,516,114 & Doxycycline                       & 1,509,230 & 0.00\% \\
  &          & Erythromycin                     & 1,507,251 & 0.00\% \\
  &          & Benzylpenicillin (penicillin G) & 1,505,830 & 0.00\% \\
\midrule
\multirow[t]{3}{=}{Pig fresh meat}
  & 17,379,368 & Doxycycline   & 789,576  & 0.02\% \\
  &          & Danofloxacin  & 785,788  & 0.00\% \\
  &          & Erythromycin  & 785,711  & 0.00\% \\
\midrule
\multirow[t]{3}{=}{Non-food animal-related matrices}
  & 7,277,222 & Diethylstilbestrol (stilbestrol) & 119,627 & 0.00\% \\
  &         & Hexestrol                        & 119,510 & 0.00\% \\
  &         & Clenbuterol                      & 115,276 & 0.02\% \\
\bottomrule
\end{tabular}
\label{table7}
\end{table*}

\begin{figure*}[htbp]
    \centering 
	\begin{minipage}{0.8\textwidth}
    \includegraphics[width=\textwidth]{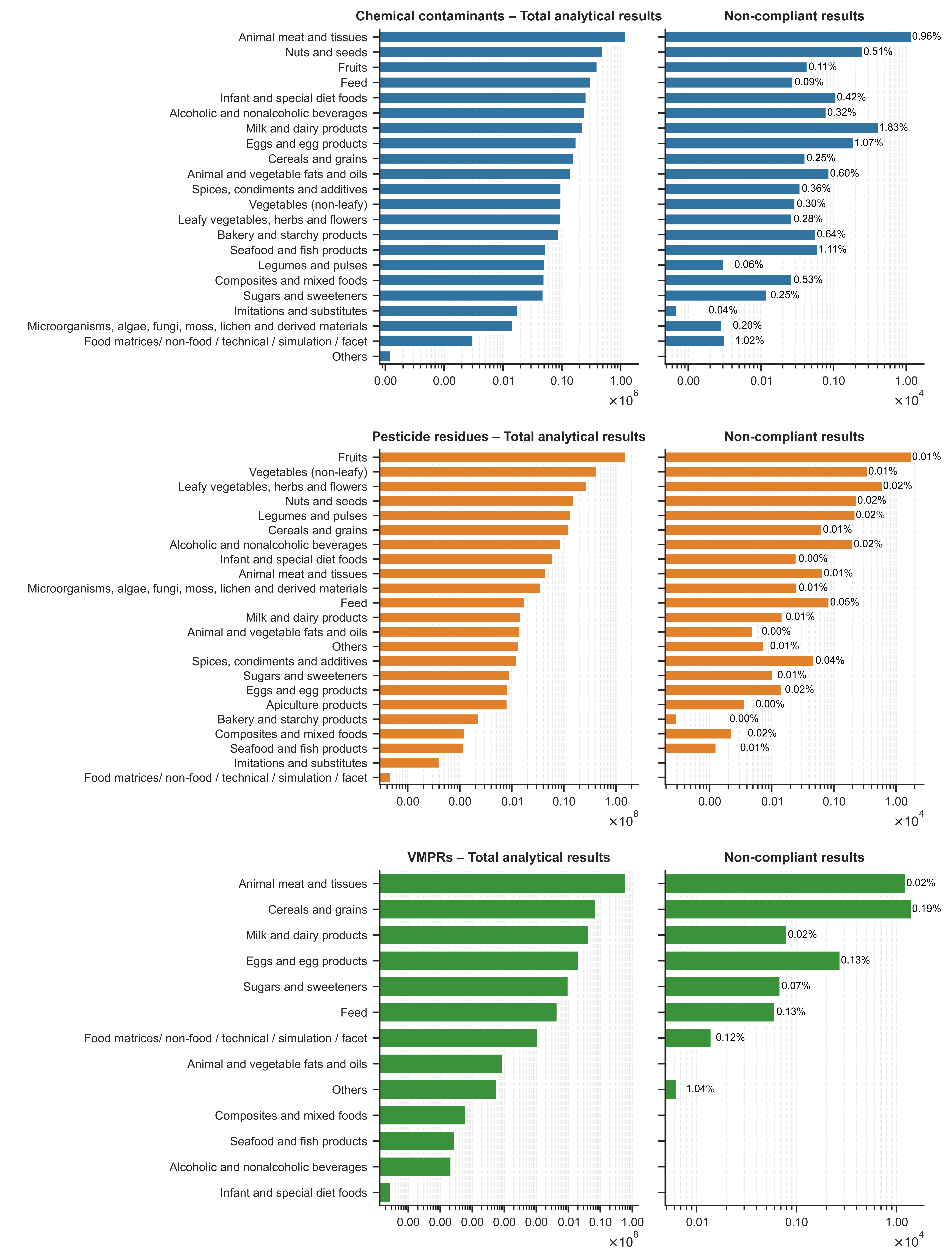}	
	\caption{Top 10 main product categories based on number of analyses per main contaminant group. Bar plots display the top 10 contaminant categories (by total analytical result count) within each of three analytical result group, chemical contaminants, pesticide residues, and veterinary medicinal product residues (VMPR). Each row contains two plots, on the left the total number of analytical results per category (log-scaled x-axis), and on the right the number of analytical results above legal limits, with percentage annotations (log-scaled x-axis). Percentage labels indicate the proportion of analytical results above legal limits relative to the total amount of analyses for each category. It should be noted that each graph uses different scales, with axes annotated using scientific notation (powers of 10), as indicated next to each axis.} 
	\label{fig5}
     \end{minipage}
    
\end{figure*}

To get a broader look, we grouped products into logical higher-classes. Among the 392,269,911 analytical results available for 23 combined food product categories, pesticide residue analytical results for “Fruit” comprised 40.6\% (159,280,616 analytical results, \autoref{fig5}. Feed had the highest percentage of analytical results above legal limits of pesticide residues (0.05\%). “Animal meat and tissues” was measured most for both VMPR (65,600,615 analytical results, 16.7\%) and chemical contaminants (1,255,344 analytical results, 0.3\%). Among the most measured food groups, the percentage of above legal limits was highest for milk and dairy products (1.83\%, \autoref{fig5}). The category “Other” had the highest percentage of analytical results above legal limits for VMPR (1.04\%) \footnote{The “Other” category largely consists of matrices grouped under "groups for hierarchies", which includes a wide range of animal and plant products that are not assigned to the core food or feed categories. These can include less frequently monitored or non-standard products such as game mammal meat and liver, other poultry meat and liver, deer and horse meat, fish and seafood, and broader groupings like "mammals and birds meat and products thereof." This pattern reflects how the data classification system uses "groups for hierarchies" as an umbrella for capturing diverse or miscellaneous products, assigning them to “Other” when they don’t fit neatly into more common or well-defined groups. In addition to the specific examples above, items such as “fruits and tree nuts”, “crops or parts of crops exclusively used for animal feed production”, and “vegetables” under pesticide residues are included in the “Other” group, since they are considered too general to be classified into more precise categories.}.

\subsection{Country-based statistics}
\begin{figure*}[htbp]
    \centering 
	\begin{minipage}{0.8\textwidth}
    \includegraphics[width=\textwidth]{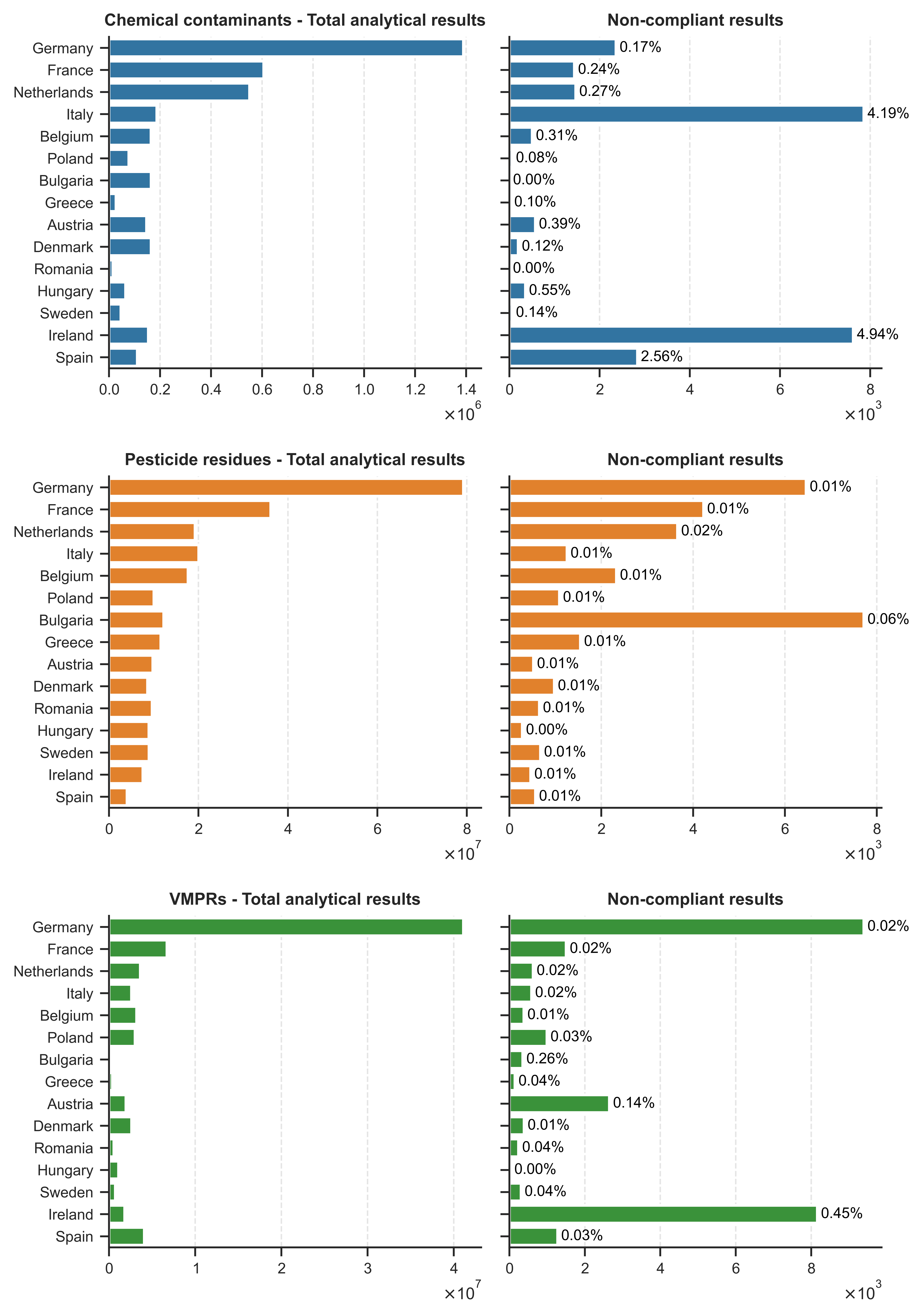}	
	\caption{Top 15 countries by total analytical results and the number of analytical results above legal limits for chemical contaminants, pesticide residues, and veterinary medicinal product residues (VMPR) analytical results, for all years available in EFSA repository. Total number of analytical results are represented on the left panel, and the number of analytical results surpassed legal limits are on the right panel. It should be noted that each graph uses different scales, with axes annotated using scientific notation (powers of 10), as indicated next to each axis.} 
	\label{fig6}
     \end{minipage}
    
\end{figure*}

The highest number of analytical results were from Germany (121,770,822; 31\%), France (43,506,669; 11\%), and the Netherlands (23,438,384; 6\%). It is important to note that these figures reflect the location of sampling, not necessarily the origin of the food products.
The different types of main contaminant groups, i.e. chemical contaminants, pesticide residues, and VMPR, were examined in \autoref{fig6}. Germany had the highest total number of analytical results for all types, with 0.17\% of the analytical results above legal limits for chemical contaminants, 0.01\% for pesticide residues, and 0.02\% VMPR. Among the top 15 countries ranked by total number of analytical results, Ireland had the highest percentage of analytical results above legal limits for chemical contaminants (4.94\%) and VMPR (0.45\%) and Bulgaria for pesticide residue analytical results (0.06\%). However, the differences in the proportion of each country’s sampling strategy per hazard type should also be noted (\autoref{figA6}). For chemical contaminants, Ireland differs from other countries with a 43.2\% convenient sampling. Germany, Belgium, and Finland stood out in pesticide residue sampling, with 81.6\% selective sampling, while for the rest of the countries objective sampling was generally performed. For VMPR, Germany’s sampling strategy was comprised of 75.9\% “other” type, whereas the rest of the countries did mostly (\textgreater{}80\%) selective sampling. 

\subsubsection{Trade flows}
\begin{figure*}[htbp]
    \centering 
    \begin{minipage}{0.8\textwidth}
	\includegraphics[width=\textwidth]{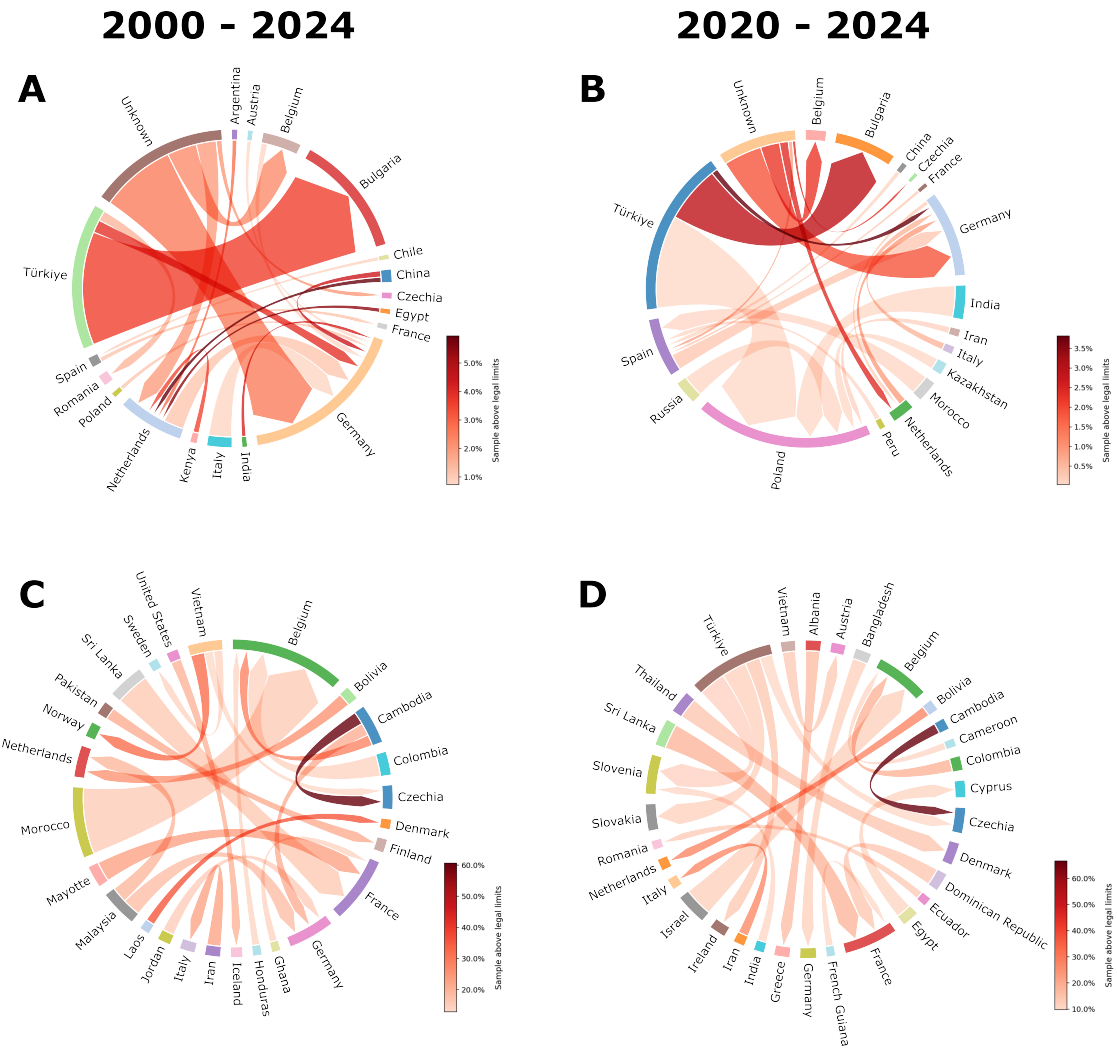}	
	\caption{Chord diagrams visualizing trade flows and samples that were measured to be above legal limits, between origin and sampling countries. The relationships between countries are shown with colored arcs, where the width of the arcs represent the total number of samples received by the sampling country and the colors represent the scaled percentage of number of samples above legal limits. Arc directions indicate the trade relationship. (A) and (C) are the origin-sampling country linkages for all available years (2002-2022) and (B) and (D) are for the last four years (2020-2022). (A) and (B) display the top 20 origin-sampling country linkages with more than 5,000 samples, ranked by the percentage of samples with analytical results above legal limits. (C) and (D) use a lower threshold of 100 samples to highlight additional high-risk linkages.} 
	\label{fig7}
    \end{minipage}
\end{figure*}

The CHEFS database also provides information on the country of origin of each monitored sample. This data can be used to gain insights in to the trade flows of food and feed contaminated above legal limits. To visualize the imported samples and the proportion of samples with analytical results above legal limits, we first excluded samples where the sampling and origin countries were the same. We then selected origin countries with more than either 5,000 samples or 100 samples in total (\autoref{fig7}).  The top 20 countries with the highest percentage of samples with hazard concentrations above legal limits were chosen for visualization in \autoref{fig7}.

Across the years 2000-2024, the origin-sampling country link with the highest percentage of samples with hazard concentrations above legal limits among links with more than 5,000 samples was China→The Netherlands. This link had 6,168 samples in total, of which 367 (5.9\%) were above legal limits (\autoref{fig7}). This was followed by Egypt→The Netherlands and China→Germany, with 5.4\% of 5,285 samples and 4.4\% of 7,527 samples with hazard concentrations above legal limits, respectively. Among the top 20 links, the origin labeled “Unknown” had the largest number of samples, 152,431 in total (\autoref{tableA5}).

This distribution changed when considering exclusively the years 2020–2024. Between the years 2000-2017, the percentage of samples with unknown origins gradually increased up to 9.8\%. However, there was a notable decline in the proportion of samples with “Unknown” origins starting in 2018, down to 0.93\% (\autoref{tableA5}). As a result, when applying the same methodology to identify the top 20 links with more than 5,000 samples for 2020–2024, a smaller proportion of samples from unknown sources appeared in the top 20 (\autoref{fig7}). Instead, 5 out of the 20 links involved samples originating from various countries but sampled in Poland, Spain, and Germany.

When a less stringent threshold was applied for the number of samples per link (n=100), and the links were ordered by the percentage of non-compliant samples, the highest percentage (60.1\%) was observed for the Cambodia→Czechia link, where 100 out of 165 (60.6\%) samples were found to be contaminated above legal limits (\autoref{fig7}). This was followed by Laos→Denmark and Vietnam→Norway, with 41 out of 116 (35.3\%) and 56 out of 176 (31.8\%) samples, respectively, found to be non-compliant with legal limits.

When considering only the most recent four years of CHEFS data (\autoref{fig7}), the highest percentage during this period was observed for the Cambodia to Czechia link, where 100 out of 150 samples (66.6\%) were non-compliant, followed by Bolivia to the Netherlands (41 out of 152 samples, 27.0\%) and Iran to Italy (39 out of 146, 26.7\%).

\section{Discussion}
\subsection{Overview}
In this paper, we provided trends from over 392 million European food safety monitoring analytical results by introducing and analyzing the CompreHensive European Food Safety (CHEFS) Database. The CHEFS database provides a unique resource for understanding trends in the occurrence of food contaminants across years, countries, and food matrices. We used the CHEFS database to demonstrate that the number of analytical results generally increased between 2000 and 2024 across the hazard categories of chemical contaminants, pesticide residues, and VMPR. A dip in the number of analytical results reported to EFSA in 2020 and 2021 most likely reflects less sampling done due to the effects of COVID-19. Another trend over the years that can be seen, is that in addition to more analytical results being reported to EFSA, the quality of the meta-data is improving. For example, the percentage of samples with an unknown origin decreased tenfold between 2017 and 2024. 

In addition to trends over years, we showed which hazards were most frequently monitored and which were most commonly found to exceed legal limits. In the chemical contaminant hazard category, heavy metals such as lead and cadmium and mycotoxins such as aflatoxin B1 were analyzed most often. The top three most analyzed pesticides were all organophosphates (chlorpyrifos, diazinon, and pirimiphos-methyl), while the top three most analyzed VMPR were antibiotics (erythromycin, danofloxacin, and doxycycline). The rates of analytical results above legal limits differed quite a lot between hazard categories with chemical contaminants having the highest non-compliant rates, followed by pesticides. VMPR had the lowest number of analytical results above legal limits even though almost only selective sampling strategies were used. This suggests that the use of veterinary drugs in food-producing animals is effectively regulated in the EU \citep{Brocca2023}. Of the foods most often analyzed, milk and dairy products were most often found to be above legal limits in the chemical contaminant group, feed was most often above legal limits for pesticides, while cereal and grains were most often above legal limits for VMPR (disregarding the ‘other’ category).

When looking at the country-based statistics, it is clear that there are big differences between the amount of analytical results between countries. It should be noted that, based on the data available, it is unclear whether this reflects more analyses done or more analytical results reported to EFSA. Furthermore, when comparing between countries, it is relevant to take into account the amount of food produced. According to the FAOSTAT data on yearly and country-based total food production, France had the highest total production at 8.45 billion tons, followed by Germany with 7.3 billion tons, and Spain with 4.37 billion tons \citep{faostat2025}. In countries like Poland and Spain, food production volumes were relatively high, while the number of analytical results remained comparatively low. However, number of analytical results did not necessarily correspond to domestic production, as food may have been sampled in a different country from its origin. Moreover, sampling intensity may be influenced by factors other than production volume, such as targeted import controls, risk-based prioritization, or specific national monitoring strategies \citep{CarrascoCabrera2023}. The analyses of the trade flows showed a trend that contaminated samples more often originated from countries outside of the EU. An important aspect to keep in mind when looking at these figures, is that even though the amount of ‘unknown’ origin has recently decreased, the country mentioned as country of origin can sometimes be the country of packing or even the port of entry \citep{Anastassiadou2019}.

In addition to the uncertainty of the accuracy of the country of origin, there are several other limitations that are important to take into account when interpreting the figures in this paper. The aforementioned differences in sampling strategies between countries and between hazard categories complicates comparisons. We showed the comparisons on the whole dataset and some subsets in the appendix. Depending on their expertise, researchers, policymakers, or other end-users might favor different comparisons. Since we publish the data and the analysis software in an open access way, those comparisons can be made by those end-users in any way that answers their specific questions. Furthermore, we did not compare levels of contaminants found with the legal limits valid at the time of analysis, but instead used the evaluation codes indicated in \autoref{tab:eval-results}. This could lead to unclarity, as now for example some analyses of vanadium or cobalt are labeled as above legal limits while there is no regulation for these metals in food and feed (Regulation (EU) 2023/915). Additionally, it is likely that EFSA will at some point open up their databases for public use phasing out the files on Zenodo \citep{EFSA2025b}. Until then the CHEFS database will be a unique and valuable infrastructure that, perhaps even more importantly, can easily be extended with other data sources.

\subsection{Future outlook}
\subsubsection{Connecting CHEFS database with external datasets}
Integrating the CHEFS database with external datasets, such as climate data, economic and geopolitical indicators, and legislative records, offers promising opportunities to investigate drivers of change indirectly influencing food safety. For example, it would enable researchers to explore associations between climate variables like humidity and temperature and the occurrence of certain toxins. Similarly, researchers could assess how quickly the prevalence of a particular contaminant changes following a regulatory adjustment to legal limits. Additionally, linking the CHEFS database to comparable food safety monitoring systems outside the EU, such as the U.S. Department of Agriculture’s Food Safety and Inspection Service (Science \& Data | Food Safety and Inspection Service, n.d.) or the Chinese National Center for Food Safety Risk Assessment (China National Center for Food Safety Risk Assessment, n.d.), would enable cross-regional comparisons and strengthen global food safety surveillance. Together, these integrations could expand the utility of the CHEFS database beyond descriptive monitoring, transforming it into a powerful tool for evaluation and prediction.

Another important source of food safety data in Europe is the Rapid Alert System for Food and Feed (RASFF), coordinated by the European Commission \citep{rasff2025}, which contains notifications on food and feed that pose a serious risk to human or animal health and is often related to incidents, such as border rejections, contamination events, or food fraud. Besides data from competent authorities, food and feed companies have a lot of in-house food safety data available. However, this data is often sensitive and it is likely not feasible to include this in the CHEFS database. An interesting option to be able to utilize this data alongside the CHEFS database is by using federated learning, which is a machine learning approach where models are trained across multiple decentralized devices or servers holding local data, without exchanging the sensitive data itself \citep{Fendor2024}.

\subsubsection{Need for increased traceability}
To meaningfully connect the CHEFS database to external datasets, it is crucial to have accurate metadata on the origin of food products. Currently, the country of origin is sometimes missing from the database, and, as mentioned above, even when reported, it is often unclear whether it refers to the place of production, processing, packaging, import or even sampling location \citep{Anastassiadou2019}. Traceability plays a key role here, as it provides detailed information about a product’s journey from farm to fork, enabling more effective risk assessments and enhancing food safety \citep{Mu2024}. For instance, knowing a more precise production location – rather than just country – and processing details can help identify the source of contamination and enable targeted interventions. In addition to improving origin information, including the date of analytical result on a more granular level (at least month) would add  great value. For analyses that integrate external factors—such as studying the impact of climate conditions on contaminants in crops—knowing the exact date, or at least the harvest or storage season, is essential for drawing reliable and actionable conclusions.

\subsubsection{Using CHEFS database with AI models}
Beyond the type of descriptive analyses reported in this paper, the large volume and richness of data in the CHEFS database open up new opportunities for advanced AI applications in food safety research and policymaking. Machine learning and other AI techniques could be used not only for forecasting the future occurrence of food safety hazards, but also for identifying patterns, and optimizing monitoring strategies. 

The use of food safety monitoring data for AI is already gaining traction. For example, \cite{Liu2021} applied Bayesian networks to develop predictive models for aflatoxin and fumonisin contamination in maize grown in Serbia. Similarly, \cite{Tarazona2022} linked monitoring data on the pesticide chlorpyrifos with biomonitoring data from European citizens to assess human exposure. \cite{Wang2023} used monitoring data to demonstrate how food safety monitoring for dioxins and dioxin-like PCBs in animal-derived products in the Netherlands could be made more cost-efficient by reallocating resources from lower-risk products (e.g., pig meat) to higher-risk ones (e.g., bovine meat) and optimizing the timing of sampling. 

Building on these examples, the CHEFS database could support the development of even more sophisticated AI models, such as deep learning approaches for anomaly detection, spatiotemporal risk prediction, or simulation of regulatory scenarios. Such applications could enhance the predictive power of food safety monitoring, support more targeted interventions, and ultimately contribute to a more proactive and efficient food safety system.

\section{Conclusion}
By integrating and harmonizing extensive monitoring data, the CompreHensive European Food Safety (CHEFS) database provides a foundation for advanced analyses and risk prediction, aiming to ultimately contribute to the safeguarding of food safety across Europe. These analyses demonstrate how the CHEFS database can serve not only as a centralized source of harmonized monitoring data, but also as a strategic tool for pinpointing priority areas in food safety policy, research, and regulation. Looking ahead, the CHEFS database could play a pivotal role in supporting EU-level rapid risk assessments during emerging food safety incidents or crises, by enabling early warning signals based on trends and historical context.

\section*{Acknowledgments}
Funding for this research has been provided by the European Union’s Horizon Europe research and innovation programmes HOLiFOOD [grant number 101059813], EFRA [grant number 101093026], and ECO-READY [grant number 101084201]. These projects have also been co-funded by the Netherlands Ministry of Agriculture, Fisheries, Food Security and Nature (LVVN) [grant numbers KB-50-005-006; grant number KB-50-005-007]. Views and opinions expressed are however those of the author(s) only and do not necessarily reflect those of the European Union or Research Executive Agency. Neither the European Union nor the granting authority can be held responsible for them. We thank Antoine Nijrolder and Gerda van Donkersgoed for our discussions.

\bibliographystyle{elsarticle-harv} 
\bibliography{references}

\begin{thebibliography}{19}
\expandafter\ifx\csname natexlab\endcsname\relax\def\natexlab#1{#1}\fi
\providecommand{\url}[1]{\texttt{#1}}
\providecommand{\href}[2]{#2}
\providecommand{\path}[1]{#1}
\providecommand{\DOIprefix}{doi:}
\providecommand{\ArXivprefix}{arXiv:}
\providecommand{\URLprefix}{URL: }
\providecommand{\Pubmedprefix}{pmid:}
\providecommand{\doi}[1]{\href{http://dx.doi.org/#1}{\path{#1}}}
\providecommand{\Pubmed}[1]{\href{pmid:#1}{\path{#1}}}
\providecommand{\bibinfo}[2]{#2}
\ifx\xfnm\relax \def\xfnm[#1]{\unskip,\space#1}\fi
\bibitem[{Anastassiadou et~al.(2019)Anastassiadou, Brancato, Brocca, Cabrera, Ferreira, Greco, Jarrah, Kazocina, Leuschner, Lostia, Magrans, Medina, Miron, Pedersen, Raczyk, Reich, Ruocco, Sacchi, Santos, Stanek, Tarazona, Theobald and Verani}]{Anastassiadou2019}
\bibinfo{author}{Anastassiadou, M.}, \bibinfo{author}{Brancato, A.}, \bibinfo{author}{Brocca, D.}, \bibinfo{author}{Cabrera, L.C.}, \bibinfo{author}{Ferreira, L.}, \bibinfo{author}{Greco, L.}, \bibinfo{author}{Jarrah, S.}, \bibinfo{author}{Kazocina, A.}, \bibinfo{author}{Leuschner, R.}, \bibinfo{author}{Lostia, A.}, \bibinfo{author}{Magrans, J.O.}, \bibinfo{author}{Medina, P.}, \bibinfo{author}{Miron, I.}, \bibinfo{author}{Pedersen, R.}, \bibinfo{author}{Raczyk, M.}, \bibinfo{author}{Reich, H.}, \bibinfo{author}{Ruocco, S.}, \bibinfo{author}{Sacchi, A.}, \bibinfo{author}{Santos, M.}, \bibinfo{author}{Stanek, A.}, \bibinfo{author}{Tarazona, J.}, \bibinfo{author}{Theobald, A.}, \bibinfo{author}{Verani, A.}, \bibinfo{year}{2019}.
\newblock \bibinfo{title}{Reporting data on pesticide residues in food and feed according to regulation (ec) no 396/2005 (2018 data collection)}.
\newblock \bibinfo{journal}{EFSA Journal} \bibinfo{volume}{17}, \bibinfo{pages}{e05655}.
\newblock \URLprefix \url{https://pmc.ncbi.nlm.nih.gov/articles/PMC7009033/}, \DOIprefix\doi{10.2903/J.EFSA.2019.5655}.
\bibitem[{Brocca and Salvatore(2023)}]{Brocca2023}
\bibinfo{author}{Brocca, D.}, \bibinfo{author}{Salvatore, S.}, \bibinfo{year}{2023}.
\newblock \bibinfo{title}{Report for 2021 on the results from the monitoring of veterinary medicinal product residues and other substances in live animals and animal products}.
\newblock \bibinfo{journal}{EFSA Supporting Publications} \bibinfo{volume}{20}.
\newblock \DOIprefix\doi{10.2903/SP.EFSA.2023.EN-7886}.
\bibitem[{Cabrera et~al.(2023)Cabrera, Piazza, Dujardin and Pastor}]{CarrascoCabrera2023}
\bibinfo{author}{Cabrera, L.C.}, \bibinfo{author}{Piazza, G.D.}, \bibinfo{author}{Dujardin, B.}, \bibinfo{author}{Pastor, P.M.}, \bibinfo{year}{2023}.
\newblock \bibinfo{title}{The 2021 european union report on pesticide residues in food}.
\newblock \bibinfo{journal}{EFSA Journal} \bibinfo{volume}{21}, \bibinfo{pages}{e07939}.
\newblock \URLprefix \url{/doi/pdf/10.2903/j.efsa.2023.7939 https://onlinelibrary.wiley.com/doi/abs/10.2903/j.efsa.2023.7939 https://efsa.onlinelibrary.wiley.com/doi/10.2903/j.efsa.2023.7939}, \DOIprefix\doi{10.2903/J.EFSA.2023.7939;PAGE:STRING:ARTICLE/CHAPTER}.
\bibitem[{EFSA(2025)}]{EFSA2025b}
\bibinfo{author}{EFSA}, \bibinfo{year}{2025}.
\newblock \bibinfo{title}{Event report symposium on data readiness for artificial intelligence symposium on data readiness for artificial intelligence} \URLprefix \url{https://efsa.onlinelibrary.wiley.com/doi/10.2903/sp.efsa.2025.EN-9434}, \DOIprefix\doi{10.2903/sp.efsa.2025.EN-9434}.
\bibitem[{EFSA et~al.(2025)EFSA, Niforou, Koffas, Livaniou and Ioannidou}]{EFSA2025}
\bibinfo{author}{EFSA}, \bibinfo{author}{Niforou, K.}, \bibinfo{author}{Koffas, N.}, \bibinfo{author}{Livaniou, A.}, \bibinfo{author}{Ioannidou, S.}, \bibinfo{year}{2025}.
\newblock \bibinfo{title}{Foodex2 maintenance 2024}.
\newblock \bibinfo{journal}{EFSA Supporting Publications} \bibinfo{volume}{22}, \bibinfo{pages}{9414E}.
\newblock \URLprefix \url{/doi/pdf/10.2903/sp.efsa.2025.EN-9414 https://onlinelibrary.wiley.com/doi/abs/10.2903/sp.efsa.2025.EN-9414 https://efsa.onlinelibrary.wiley.com/doi/10.2903/sp.efsa.2025.EN-9414}, \DOIprefix\doi{10.2903/SP.EFSA.2025.EN-9414}.
\bibitem[{{European Food Safety Authority}(2013)}]{EFSA2013}
\bibinfo{author}{{European Food Safety Authority}}, \bibinfo{year}{2013}.
\newblock \bibinfo{title}{Standard sample description ver. 2.0}.
\newblock \bibinfo{journal}{EFSA Journal} \bibinfo{volume}{11}, \bibinfo{pages}{3424}.
\newblock \URLprefix \url{https://efsa.onlinelibrary.wiley.com/doi/10.2903/j.efsa.2013.3424}, \DOIprefix\doi{10.2903/j.efsa.2013.3424}.
\bibitem[{Fanzo et~al.(2021)Fanzo, Bellows, Spiker, Thorne-Lyman and Bloem}]{Fanzo2021}
\bibinfo{author}{Fanzo, J.}, \bibinfo{author}{Bellows, A.L.}, \bibinfo{author}{Spiker, M.L.}, \bibinfo{author}{Thorne-Lyman, A.L.}, \bibinfo{author}{Bloem, M.W.}, \bibinfo{year}{2021}.
\newblock \bibinfo{title}{The importance of food systems and the environment for nutrition}.
\newblock \bibinfo{journal}{The American Journal of Clinical Nutrition} \bibinfo{volume}{113}, \bibinfo{pages}{7--16}.
\newblock \URLprefix \url{https://www.sciencedirect.com/science/article/pii/S0002916522005536}, \DOIprefix\doi{10.1093/AJCN/NQAA313}.
\bibitem[{{FAOSTAT}(n.d.)}]{faostat2025}
\bibinfo{author}{{FAOSTAT}}, \bibinfo{year}{n.d.}
\newblock \bibinfo{title}{Faostat}.
\newblock \bibinfo{howpublished}{\url{https://www.fao.org/faostat/en/}}.
\newblock \bibinfo{note}{Retrieved May 8, 2025}.
\bibitem[{van~der Fels-Klerx et~al.(2024)van~der Fels-Klerx, van Asselt, van Leeuwen, Dorgelo and van~den Hil}]{vanderFels-Klerx2024}
\bibinfo{author}{van~der Fels-Klerx, H.J.}, \bibinfo{author}{van Asselt, E.D.}, \bibinfo{author}{van Leeuwen, S.P.}, \bibinfo{author}{Dorgelo, F.O.}, \bibinfo{author}{van~den Hil, E.F.H.}, \bibinfo{year}{2024}.
\newblock \bibinfo{title}{Prioritization of chemical food safety hazards in the european feed supply chain}.
\newblock \bibinfo{journal}{Comprehensive Reviews in Food Science and Food Safety} \bibinfo{volume}{23}, \bibinfo{pages}{e70025}.
\newblock \URLprefix \url{/doi/pdf/10.1111/1541-4337.70025 https://onlinelibrary.wiley.com/doi/abs/10.1111/1541-4337.70025 https://ift.onlinelibrary.wiley.com/doi/10.1111/1541-4337.70025}, \DOIprefix\doi{10.1111/1541-4337.70025;SUBPAGE:STRING:FULL}.
\bibitem[{Fendor et~al.(2024)Fendor, van~der Velden, Wang, Carnoli, Mutlu and Hürriyetoğlu}]{Fendor2024}
\bibinfo{author}{Fendor, Z.}, \bibinfo{author}{van~der Velden, B.H.M.}, \bibinfo{author}{Wang, X.}, \bibinfo{author}{Carnoli, A.J.}, \bibinfo{author}{Mutlu, O.}, \bibinfo{author}{Hürriyetoğlu, A.}, \bibinfo{year}{2024}.
\newblock \bibinfo{title}{Federated learning in food research}.
\newblock \URLprefix \url{http://arxiv.org/abs/2406.06202}, \DOIprefix\doi{10.48550/arXiv.2406.06202}. \bibinfo{note}{arXiv:2406.06202 [cs]}.
\bibitem[{Focker et~al.(2018)Focker, van~der Fels-Klerx and Lansink}]{Focker2018}
\bibinfo{author}{Focker, M.M.}, \bibinfo{author}{van~der Fels-Klerx, H.J.}, \bibinfo{author}{Lansink, A.G.O.}, \bibinfo{year}{2018}.
\newblock \bibinfo{title}{Systematic review of methods to determine the cost-effectiveness of monitoring plans for chemical and biological hazards in the life sciences}.
\newblock \bibinfo{journal}{Comprehensive Reviews in Food Science and Food Safety} \bibinfo{volume}{17}, \bibinfo{pages}{633--645}.
\newblock \URLprefix \url{https://onlinelibrary.wiley.com/doi/full/10.1111/1541-4337.12340 https://onlinelibrary.wiley.com/doi/abs/10.1111/1541-4337.12340 https://ift.onlinelibrary.wiley.com/doi/10.1111/1541-4337.12340}, \DOIprefix\doi{10.1111/1541-4337.12340}.
\bibitem[{Gruber(1993)}]{Gruber1993}
\bibinfo{author}{Gruber, T.R.}, \bibinfo{year}{1993}.
\newblock \bibinfo{title}{A translation approach to portable ontology specifications}.
\newblock \bibinfo{journal}{Knowledge Acquisition} \bibinfo{volume}{5}, \bibinfo{pages}{199--220}.
\newblock \URLprefix \url{https://www.sciencedirect.com/science/article/pii/S1042814383710083}, \DOIprefix\doi{10.1006/KNAC.1993.1008}.
\bibitem[{Hu et~al.(2023)Hu, Ma, Wang, Zhang, Li and Yu}]{Hu2023}
\bibinfo{author}{Hu, Y.}, \bibinfo{author}{Ma, B.}, \bibinfo{author}{Wang, H.}, \bibinfo{author}{Zhang, Y.}, \bibinfo{author}{Li, Y.}, \bibinfo{author}{Yu, G.}, \bibinfo{year}{2023}.
\newblock \bibinfo{title}{Detecting different pesticide residues on hami melon surface using hyperspectral imaging combined with 1d-cnn and information fusion}.
\newblock \bibinfo{journal}{Frontiers in plant science} \bibinfo{volume}{14}, \bibinfo{pages}{1105601}.
\newblock \URLprefix \url{https://pubmed.ncbi.nlm.nih.gov/37223822}, \DOIprefix\doi{10.3389/fpls.2023.1105601}. \bibinfo{note}{place: Switzerland}.
\bibitem[{Liu et~al.(2021)Liu, Liu, Dudas, Loc, Bagi and van~der Fels-Klerx}]{Liu2021}
\bibinfo{author}{Liu, N.}, \bibinfo{author}{Liu, C.}, \bibinfo{author}{Dudas, T.N.}, \bibinfo{author}{Loc, M.C.}, \bibinfo{author}{Bagi, F.F.}, \bibinfo{author}{van~der Fels-Klerx, H.J.}, \bibinfo{year}{2021}.
\newblock \bibinfo{title}{Improved aflatoxins and fumonisins forecasting models for maize (prema and prefum), using combined mechanistic and bayesian network modeling-serbia as a case study}.
\newblock \bibinfo{journal}{Frontiers in microbiology} \bibinfo{volume}{12}, \bibinfo{pages}{643604}.
\newblock \URLprefix \url{https://pubmed.ncbi.nlm.nih.gov/33967981}, \DOIprefix\doi{10.3389/fmicb.2021.643604}. \bibinfo{note}{place: Switzerland}.
\bibitem[{Mu et~al.(2024)Mu, Kleter, Bouzembrak, Dupouy, Frewer, Natour and Marvin}]{Mu2024}
\bibinfo{author}{Mu, W.}, \bibinfo{author}{Kleter, G.A.}, \bibinfo{author}{Bouzembrak, Y.}, \bibinfo{author}{Dupouy, E.}, \bibinfo{author}{Frewer, L.J.}, \bibinfo{author}{Natour, F.N.R.A.}, \bibinfo{author}{Marvin, H.J.}, \bibinfo{year}{2024}.
\newblock \bibinfo{title}{Making food systems more resilient to food safety risks by including artificial intelligence, big data, and internet of things into food safety early warning and emerging risk identification tools}.
\newblock \bibinfo{journal}{Comprehensive Reviews in Food Science and Food Safety} \bibinfo{volume}{23}, \bibinfo{pages}{1--18}.
\newblock \URLprefix \url{/doi/pdf/10.1111/1541-4337.13296 https://onlinelibrary.wiley.com/doi/abs/10.1111/1541-4337.13296 https://ift.onlinelibrary.wiley.com/doi/10.1111/1541-4337.13296}, \DOIprefix\doi{10.1111/1541-4337.13296;SUBPAGE:STRING:FULL}.
\bibitem[{{RASFF - European Commission}(n.d.)}]{rasff2025}
\bibinfo{author}{{RASFF - European Commission}}, \bibinfo{year}{n.d.}
\newblock \bibinfo{title}{Rasff - rapid alert system for food and feed}.
\newblock \bibinfo{howpublished}{\url{https://food.ec.europa.eu/food-safety/rasff_en}}.
\newblock \bibinfo{note}{Retrieved July 11, 2025}.
\bibitem[{Sorbo et~al.(2022)Sorbo, Pucci, Nobili, Taglieri, Passeri and Zoani}]{Sorbo2022}
\bibinfo{author}{Sorbo, A.}, \bibinfo{author}{Pucci, E.}, \bibinfo{author}{Nobili, C.}, \bibinfo{author}{Taglieri, I.}, \bibinfo{author}{Passeri, D.}, \bibinfo{author}{Zoani, C.}, \bibinfo{year}{2022}.
\newblock \bibinfo{title}{Food safety assessment: Overview of metrological issues and regulatory aspects in the european union}.
\newblock \bibinfo{journal}{Separations} \bibinfo{volume}{9}, \bibinfo{pages}{53}.
\newblock \URLprefix \url{https://www.mdpi.com/2297-8739/9/2/53}, \DOIprefix\doi{10.3390/separations9020053}. \bibinfo{note}{number: 2 Publisher: Multidisciplinary Digital Publishing Institute}.
\bibitem[{Tarazona et~al.(2022)Tarazona, González-Caballero, de~Alba-Gonzalez, Pedraza-Diaz, Cañas, Dominguez-Morueco, Esteban-López, Cattaneo, Katsonouri, Makris, Halldorsson, Olafsdottir, Zock, Dias, Decker, Morrens, Berman, Barnett-Itzhaki, Lindh, Gilles, Govarts, Schoeters, Weber, Kolossa-Gehring, Santonen and Castaño}]{Tarazona2022}
\bibinfo{author}{Tarazona, J.V.}, \bibinfo{author}{González-Caballero, M.D.C.}, \bibinfo{author}{de~Alba-Gonzalez, M.}, \bibinfo{author}{Pedraza-Diaz, S.}, \bibinfo{author}{Cañas, A.}, \bibinfo{author}{Dominguez-Morueco, N.}, \bibinfo{author}{Esteban-López, M.}, \bibinfo{author}{Cattaneo, I.}, \bibinfo{author}{Katsonouri, A.}, \bibinfo{author}{Makris, K.C.}, \bibinfo{author}{Halldorsson, T.I.}, \bibinfo{author}{Olafsdottir, K.}, \bibinfo{author}{Zock, J.P.}, \bibinfo{author}{Dias, J.}, \bibinfo{author}{Decker, A.D.}, \bibinfo{author}{Morrens, B.}, \bibinfo{author}{Berman, T.}, \bibinfo{author}{Barnett-Itzhaki, Z.}, \bibinfo{author}{Lindh, C.}, \bibinfo{author}{Gilles, L.}, \bibinfo{author}{Govarts, E.}, \bibinfo{author}{Schoeters, G.}, \bibinfo{author}{Weber, T.}, \bibinfo{author}{Kolossa-Gehring, M.}, \bibinfo{author}{Santonen, T.}, \bibinfo{author}{Castaño, A.}, \bibinfo{year}{2022}.
\newblock \bibinfo{title}{Improving the risk assessment of pesticides through the integration of human biomonitoring and food monitoring data: A case study for chlorpyrifos}.
\newblock \bibinfo{journal}{Toxics} \bibinfo{volume}{10}, \bibinfo{pages}{313}.
\newblock \URLprefix \url{https://www.mdpi.com/2305-6304/10/6/313/htm https://www.mdpi.com/2305-6304/10/6/313}, \DOIprefix\doi{10.3390/TOXICS10060313/S1}.
\bibitem[{Wang et~al.(2023)Wang, van~der Fels-Klerx and Lansink}]{Wang2023}
\bibinfo{author}{Wang, Z.}, \bibinfo{author}{van~der Fels-Klerx, H.J.}, \bibinfo{author}{Lansink, A.G.O.}, \bibinfo{year}{2023}.
\newblock \bibinfo{title}{Designing optimal food safety monitoring schemes using bayesian network and integer programming: The case of monitoring dioxins and dl-pcbs}.
\newblock \bibinfo{journal}{Risk Analysis} \bibinfo{volume}{43}, \bibinfo{pages}{1400--1413}.
\newblock \URLprefix \url{https://onlinelibrary.wiley.com/doi/full/10.1111/risa.14030 https://onlinelibrary.wiley.com/doi/abs/10.1111/risa.14030 https://onlinelibrary.wiley.com/doi/10.1111/risa.14030}, \DOIprefix\doi{10.1111/RISA.14030}.

\end{thebibliography}

\clearpage
\onecolumn
\appendix
\setcounter{table}{0}
\setcounter{figure}{0}
\renewcommand\thesection{\Alph{section}}
\section{Appendix}

\textbf{Supplemental figures}

Figure A1: Design of the database

Figure A2: Missing values in the chemical contaminant datasets

Figure A3: Relationship between total analytical results and the number of samples

Figure A4: Shared contaminant names across hazard types

Figure A5: Second level ontologies of contaminant groups

Figure A6: Distribution of sampling strategies by countries for each hazard type

\clearpage
\newpage

\begin{figure*}[htbp]
    \centering 
    \includegraphics[width=\textwidth]{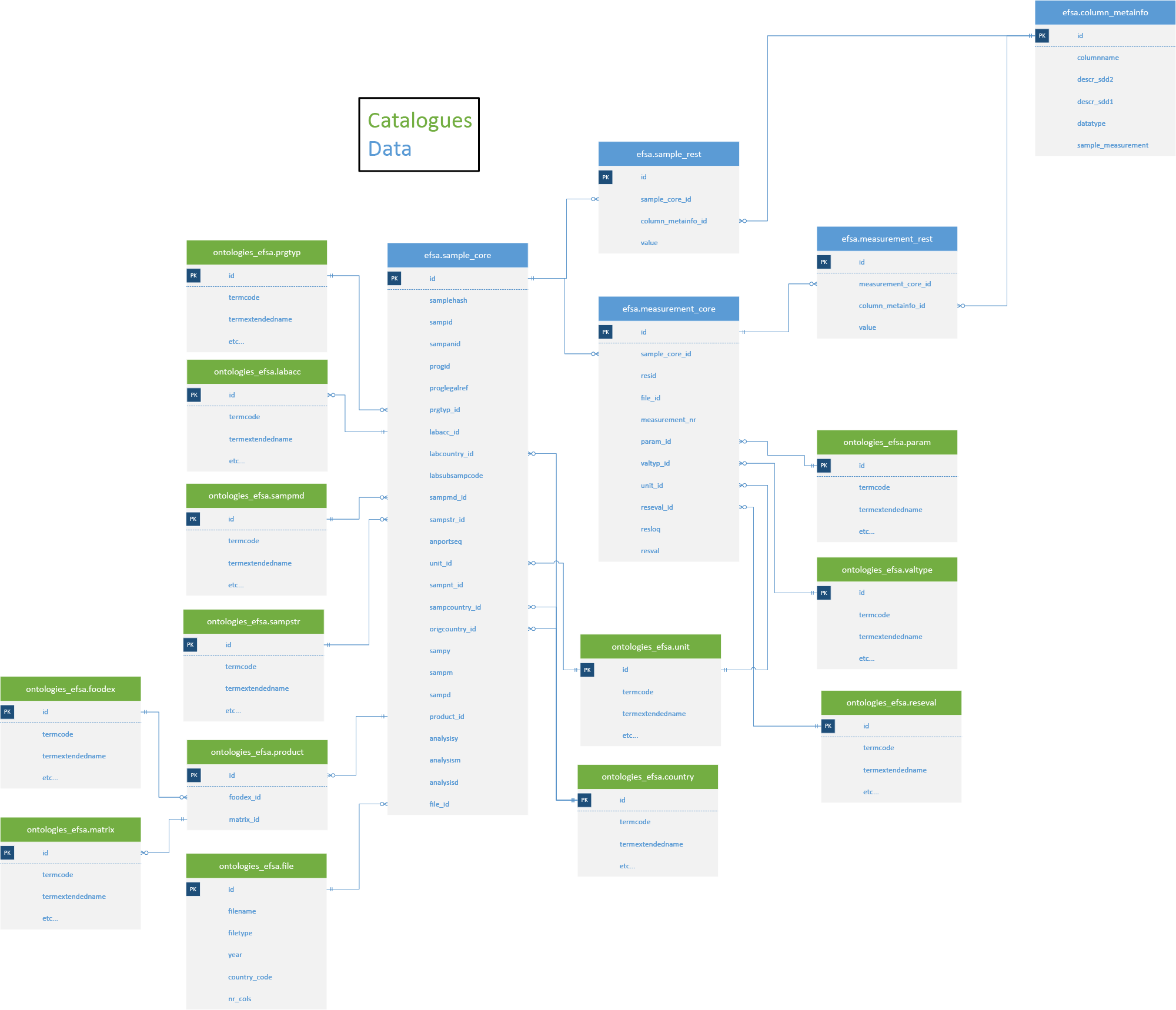}	
	\caption{Design of the database. The blue tables store the sample and analytical result information, plus meta-data on each variable.} 
	\label{figA1}
\end{figure*}

\begin{figure*}[htbp]
    \centering 
    \includegraphics[width=\textwidth]{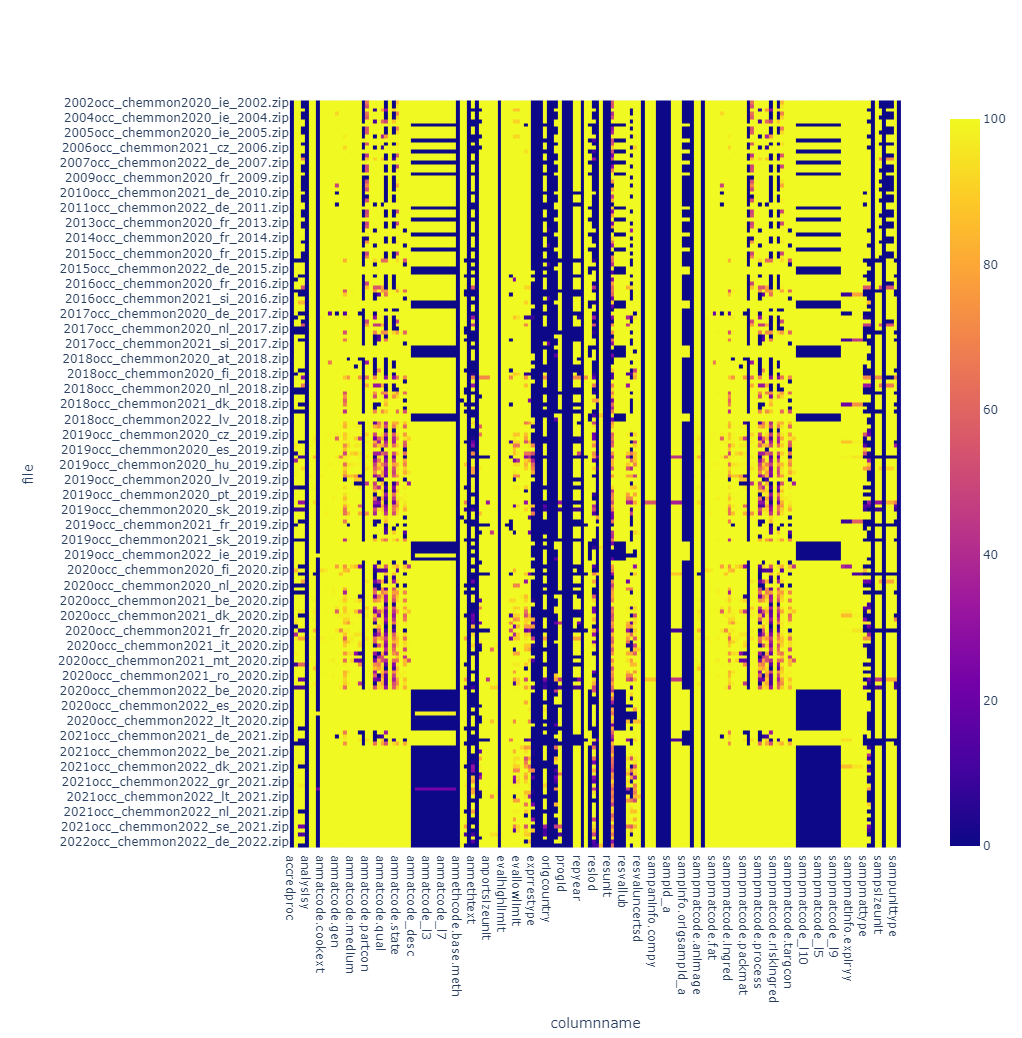}	
	\caption{Missing values in the chemical contaminant datasets. On the Y-axis are the data files, and on the X-axis the variables. Data sparsity can be due to many variables that are not mandatory to report to EFSA, leading to many missing data in those variables. A yellow vertical bar means that a variable has mostly missing values across all files. A blue vertical bar means that a variable has little to no missing values across all files. A vertical bar with both yellow and blue colors represents a variable that in some files has a lot of missing values and in other files has little to no missing values.} 
	\label{figA2}
\end{figure*}

\begin{figure*}[htbp]
    \centering 
    \includegraphics[width=\textwidth]{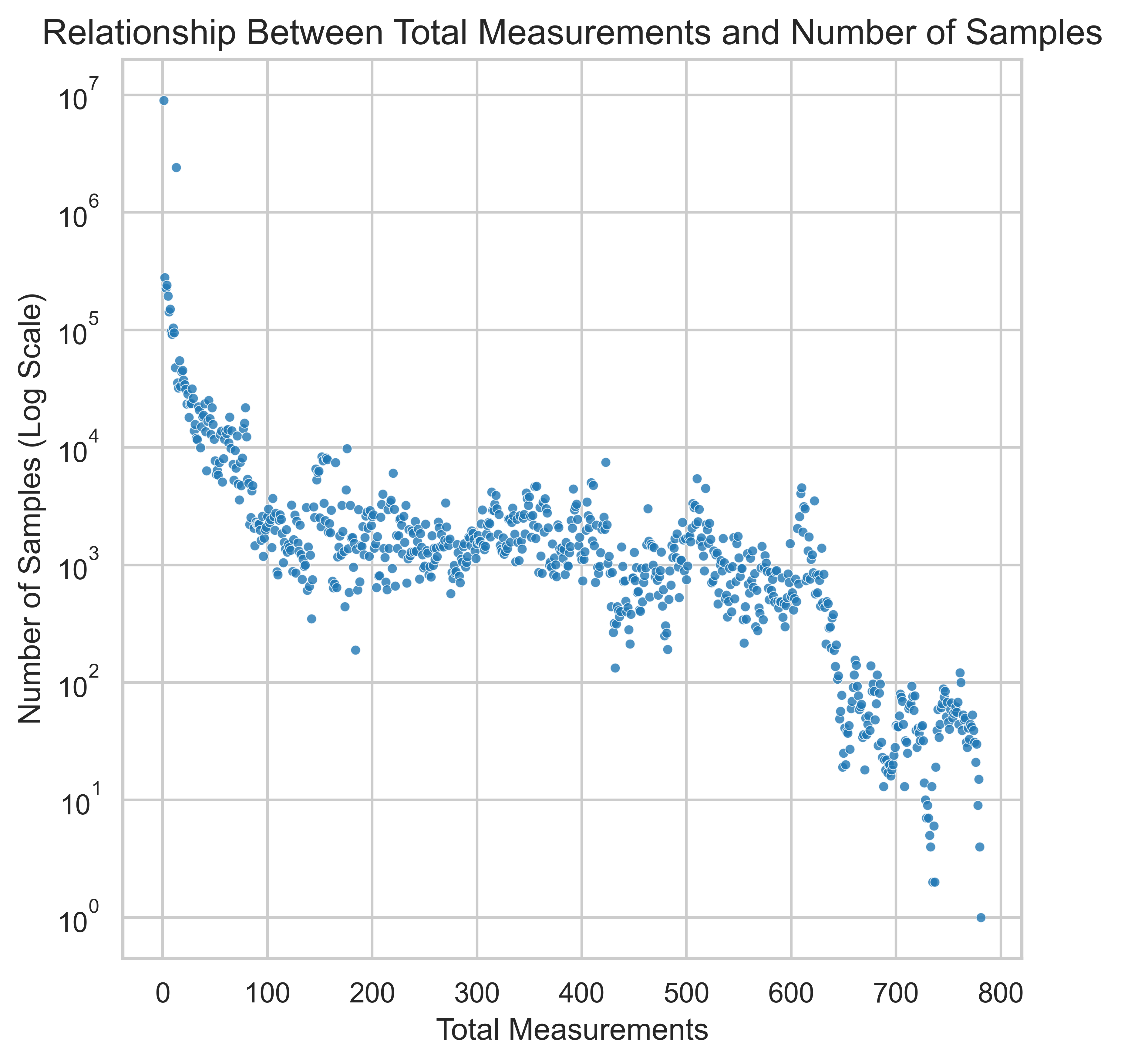}	
	\caption{Relationship between total analytical results and the number of samples available in CHEFS data. The x-axis represents the total number of results, while the y-axis shows the number of samples on a log scale.} 
	\label{figA3}
\end{figure*}

\begin{figure*}[htbp]
    \centering 
    \includegraphics[width=\textwidth]{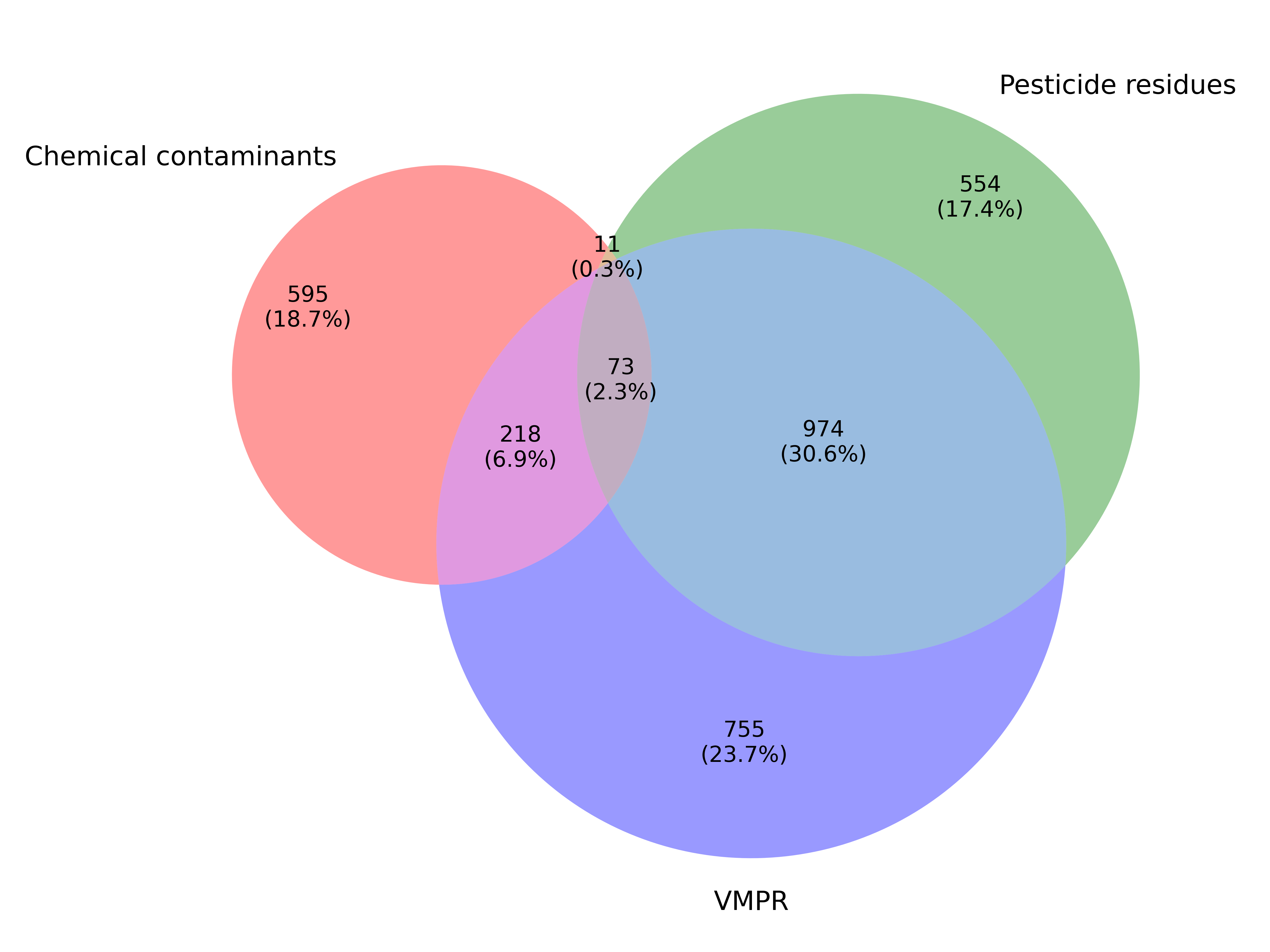}	
	\caption{Shared contaminant names across hazard types.} 
	\label{figA4}
\end{figure*}

\begin{figure*}[htbp]
    \centering 
    \includegraphics[width=.8\textwidth]{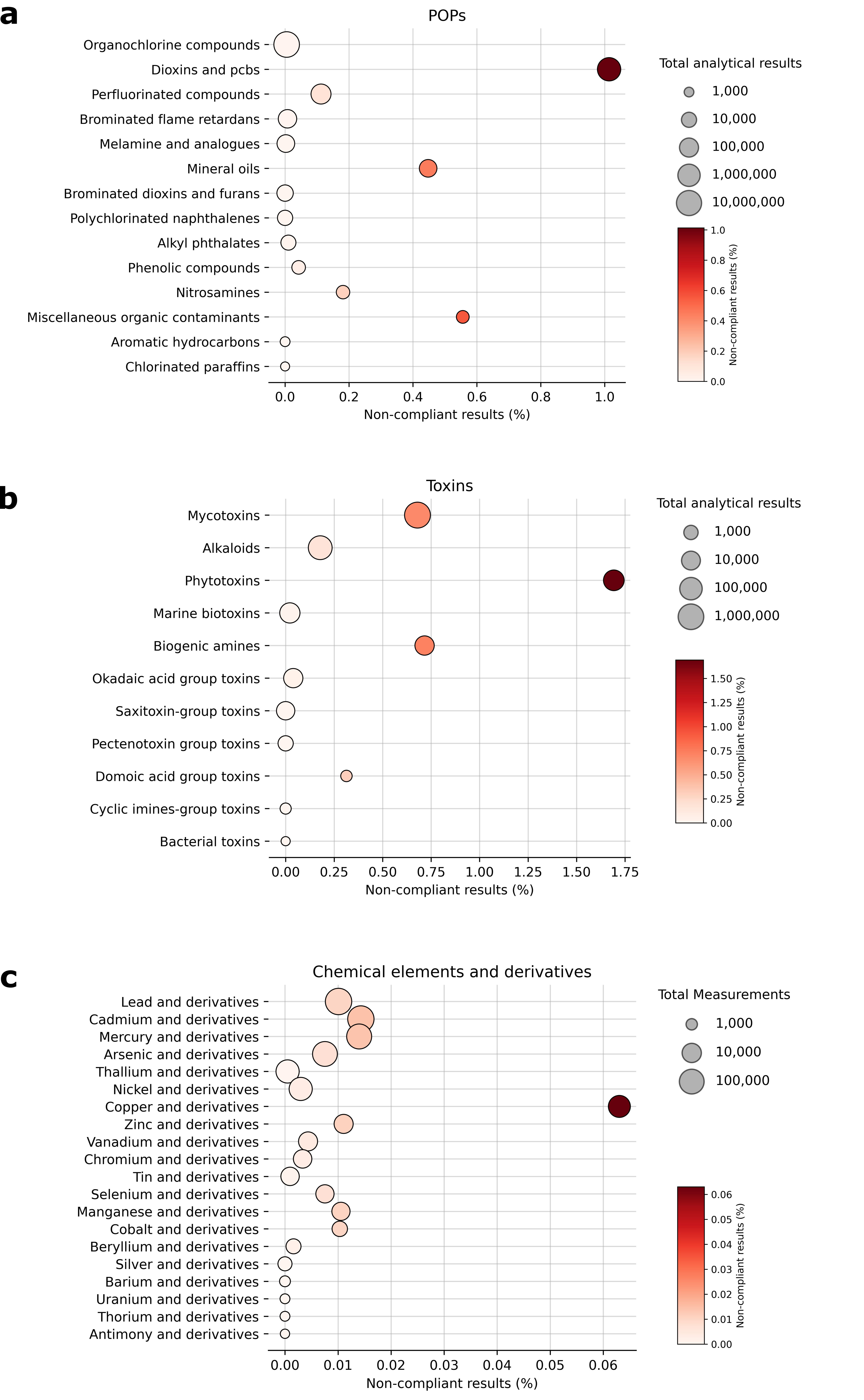}	
	\caption{Second level ontologies of contaminant groups and total number of analytical results vs non-compliance ratios for a) “persistent organic pollutants (POPs) and other organic contaminants”, b) “toxins”, and c) “chemical elements (including derivatives) and others”. The size of bubbles indicate the number of total number of analytical results whereas the color indicate the proportion of non-compliance.} 
	\label{figA5}
\end{figure*}

\begin{figure*}[htbp]
    \centering 
    \includegraphics[width=\textwidth]{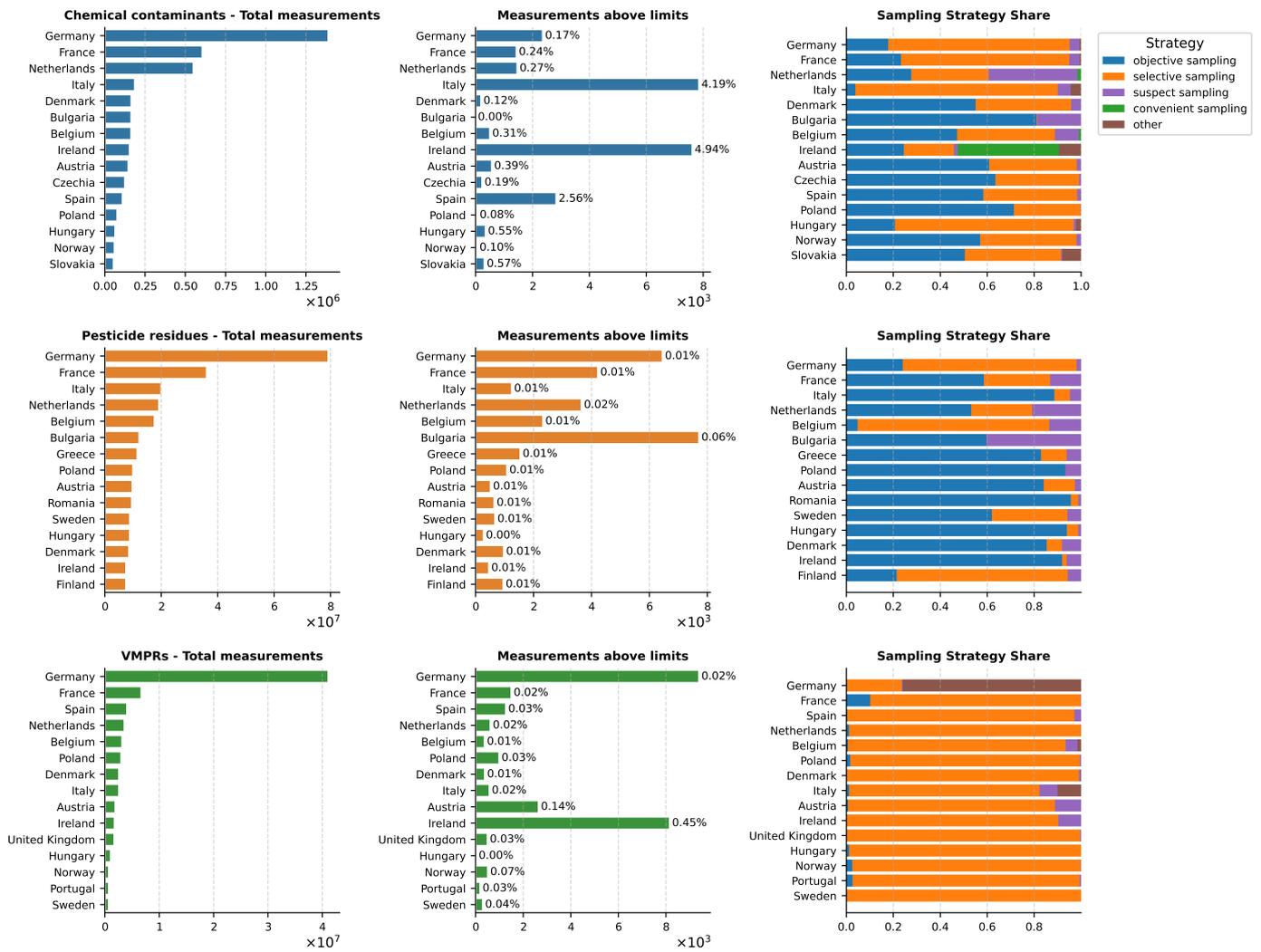}	
	\caption{Distribution of sampling strategies by countries for each hazard type.} 
	\label{figA6}
\end{figure*}

\newpage

\textbf{Supplemental tables}

Table A1: Grouping dictionary for product names.

Table A2: Numbers of analytical results and samples per hazard type.

Table A3: Sample number distribution over sampling strategy. 

Table A4: Distribution of number of analytical results across different sampling strategies between 2015 - 2019

Table A5: Proportion of samples with unknown country origins over years.

\clearpage

\footnotesize
\begin{longtable}{p{8cm}p{8cm}}

\caption{Grouping dictionary for product names.}
\label{tableA1} \\
\toprule
\textbf{Category} & \textbf{Item} \\
\midrule
\endfirsthead

\toprule
\textbf{Category} & \textbf{Item} \\
\midrule
\endhead

\midrule
\multicolumn{2}{r}{\textit{Continues on next page}} \\
\midrule
\endfoot

\bottomrule
\endlastfoot

Alcoholic and Nonalcoholic Beverages & alcoholic beverages \\
Alcoholic and Nonalcoholic Beverages & fruit / vegetable juices and nectars \\
Alcoholic and Nonalcoholic Beverages & hot drinks and similar (coffee, cocoa, tea and herbal infusions) \\
Alcoholic and Nonalcoholic Beverages & tea, coffee, herbal infusions, cocoa and carobs \\
Alcoholic and Nonalcoholic Beverages & water, water-based beverages and related ingredients \\
Alcoholic and Nonalcoholic Beverages & liquid or gel separated from plant rpcs \\
Alcoholic and Nonalcoholic Beverages & derivatives of coffee, cocoa, tea, herbal infusion materials and similar rpcs \\
Alcoholic and Nonalcoholic Beverages & hot drinks and infusions composite ingredients \\
Alcoholic and Nonalcoholic Beverages & materials for infusions or hot drinks of miscellaneous origin \\
animal and vegetable fats and oils & animal and vegetable fats and oils and primary derivatives thereof \\
Animal Meat and Tissues & animal carcase \\
Animal Meat and Tissues & animal fresh fat tissues \\
Animal Meat and Tissues & animal fresh meat \\
Animal Meat and Tissues & animal meat commodities (tissue rpcs) \\
Animal Meat and Tissues & animal mechanically separated meat (msm) \\
Animal Meat and Tissues & animal offal and other slaughtering products \\
Animal Meat and Tissues & animal tissues \\
Animal Meat and Tissues & bovine meat \\
Animal Meat and Tissues & canned-tinned meat \\
Animal Meat and Tissues & charcuterie meat products \\
Animal Meat and Tissues & goat meat \\
Animal Meat and Tissues & horses, asses, mules or hinnies meat \\
Animal Meat and Tissues & marinated meat \\
Animal Meat and Tissues & other farm animals meat \\
Animal Meat and Tissues & poultry meat \\
Animal Meat and Tissues & sheep meat \\
Animal Meat and Tissues & swine meat \\
Animal Meat and Tissues & terrestrial invertebrate animals \\
Animal Meat and Tissues & wild terrestrial vertebrate animals \\
Animal Meat and Tissues & Other processed or preserved meat \\
Animal Meat and Tissues & amphibians and reptiles \\
Animal Meat and Tissues & Other animals (as animal) \\
Apiculture Products & honey and other apicultural products \\
Bakery and Starchy Products & bakery products \\
Bakery and Starchy Products & pasta and similar products \\
Bakery and Starchy Products & raw doughs and pre-mixes \\
Bakery and Starchy Products & starches \\
Others & Other fruits and nuts, vegetables and other plant products \\
Others & Other groups for different domains but not all \\
Others & Other groups for hierarchies \\
Others & Other groups for pesticides \\
Others & other products \\
Others & Other VMPR \\
Cereals and Grains & barley \\
Cereals and Grains & buckwheat and other pseudo-cereals \\
Cereals and Grains & cereal primary derivatives \\
Cereals and Grains & cereals \\
Cereals and Grains & common millet/proso millet \\
Cereals and Grains & maize/corn \\
Cereals and Grains & oat \\
Cereals and Grains & other cereals \\
Cereals and Grains & rice \\
Cereals and Grains & rye \\
Cereals and Grains & sorghum \\
Cereals and Grains & wheat \\
Composites and Mixed Foods & cereals- or roots-based snacks or breakfast composite \\
Composites and Mixed Foods & composite dishes \\
Composites and Mixed Foods & confectionery including chocolate \\
Eggs and Egg products & birds eggs \\
Eggs and Egg products & egg powder \\
Eggs and Egg products & hardened egg products \\
Eggs and Egg products & liquid egg products \\
Eggs and Egg products & whole eggs \\
Feed & feed \\
Food Matrices/ Non-Food / Technical / Simulation / Facet & facets \\
Food Matrices/ Non-Food / Technical / Simulation / Facet & Other non-food matrices \\
Food Matrices/ Non-Food / Technical / Simulation / Facet & ingredients for food fortification/enrichment and supplements \\
Food Matrices/ Non-Food / Technical / Simulation / Facet & non-food matrices \\
Food Matrices/ Non-Food / Technical / Simulation / Facet & Other environment \\
Food Matrices/ Non-Food / Technical / Simulation / Facet & Other food contact materials \\
Food Matrices/ Non-Food / Technical / Simulation / Facet & Other food simulants \\
Food Matrices/ Non-Food / Technical / Simulation / Facet & Other non-food animal-related matrices \\
Fruits & berries and small fruits \\
Fruits & citrus fruits \\
Fruits & fruit rpcs \\
Fruits & fruit/vegetable juice concentrate \\
Fruits & fruit/vegetable juice powder \\
Fruits & fruiting vegetables \\
Fruits & miscellaneous fruits \\
Fruits & pome fruits \\
Fruits & processed or preserved fruits \\
Fruits & stone fruits \\
Fruits & fruit / vegetable spreads and similar \\
Imitations and Substitutes & isolated proteins and other protein products \\
Imitations and Substitutes & meat and dairy imitates \\
Infant and Special Diet Foods & food for infants and young children \\
Infant and Special Diet Foods & food for particular diets \\
Infant and Special Diet Foods & food products for young population \\
Infant and Special Diet Foods & infant formulae/follow-on formulae \\
Leafy Vegetables, Herbs and Flowers & flowering brassica \\
Leafy Vegetables, Herbs and Flowers & flowers \\
Leafy Vegetables, Herbs and Flowers & hops, dried \\
Leafy Vegetables, Herbs and Flowers & leaf vegetables, herbs and edible flowers \\
Leafy Vegetables, Herbs and Flowers & leaves \\
Legumes and Pulses & legume vegetables \\
Legumes and Pulses & legumes with pod \\
Legumes and Pulses & processed or preserved legumes \\
Legumes and Pulses & pulses (dry) \\
Microorganisms, Algae, Fungi, Moss, Lichen and Derived Materials & algae and prokaryotes organisms \\
Microorganisms, Algae, Fungi, Moss, Lichen and Derived Materials & fungi \\
Microorganisms, Algae, Fungi, Moss, Lichen and Derived Materials & fungi, mosses and lichens \\
Microorganisms, Algae, Fungi, Moss, Lichen and Derived Materials & plants where the vegetative tissue is used as food \\
Microorganisms, Algae, Fungi, Moss, Lichen and Derived Materials & microbiological or enzymatic ingredients \\
Milk and Dairy Products & baked milk and similar \\
Milk and Dairy Products & buttermilk \\
Milk and Dairy Products & cheese \\
Milk and Dairy Products & cream and cream products \\
Milk and Dairy Products & dairy snacks \\
Milk and Dairy Products & fermented milk products \\
Milk and Dairy Products & flavoured milks \\
Milk and Dairy Products & milk \\
Milk and Dairy Products & milk and dairy powders and concentrates \\
Milk and Dairy Products & sour cream products \\
Milk and Dairy Products & whey \\
Nuts and Seeds & oilseeds and oil fruits \\
Nuts and Seeds & primary derivatives from nuts and similar seeds \\
Nuts and Seeds & seeds \\
Nuts and Seeds & tree nuts \\
Seafood and Fish Products & canned seafood \\
Seafood and Fish Products & canned/jarred fish \\
Seafood and Fish Products & dried fish \\
Seafood and Fish Products & dried seafood \\
Seafood and Fish Products & marinated / pickled fish \\
Seafood and Fish Products & marinated / pickled seafood \\
Seafood and Fish Products & Other processed or preserved fish (including processed offal) \\
Seafood and Fish Products & Other processed or preserved seafood \\
Seafood and Fish Products & products of animal origin - fish, fish products and any other marine and freshwater food products \\
Seafood and Fish Products & salted seafood \\
Seafood and Fish Products & salt-preserved fish \\
Seafood and Fish Products & smoked fish \\
Seafood and Fish Products & smoked seafood \\
Seafood and Fish Products & structured/textured fish meat products or fish paste \\
Spices, Condiments and Additives & aril spices \\
Spices, Condiments and Additives & bark spices \\
Spices, Condiments and Additives & bud spices \\
Spices, Condiments and Additives & extracts of plant origin \\
Spices, Condiments and Additives & major isolated ingredients, additives, flavours, baking and processing aids \\
Spices, Condiments and Additives & seasoning, sauces and condiments \\
Spices, Condiments and Additives & spices, dried \\
Spices, Condiments and Additives & processed or preserved herbs, spices and similar \\
Sugars and Sweeteners & sugars and similar \\
Sugars and Sweeteners & table-top sweeteners formulations \\
Vegetables (Non-Leafy) & brassica vegetables (excluding brassica roots and brassica baby leaf crops) \\
Vegetables (Non-Leafy) & bulb vegetables \\
Vegetables (Non-Leafy) & processed or preserved vegetables and similar \\
Vegetables (Non-Leafy) & root and tuber vegetables \\
Vegetables (Non-Leafy) & roots and other underground parts \\
Vegetables (Non-Leafy) & sprouts, shoots and similar \\
Vegetables (Non-Leafy) & stem vegetables \\
Vegetables (Non-Leafy) & stems/stalks \\
Vegetables (Non-Leafy) & sugar plants \\
\end{longtable}

\clearpage
\newpage

\begin{table}[htbp]
\caption{Numbers of analytical results and samples per hazard type. VMPR = Veterinary medical product residues.}
\label{tableA2}
\centering
\footnotesize

\begin{tabular}{p{3 cm} R{3cm} R{3cm} R{3cm}}
\toprule
& \textbf{Chemical contaminants} & \textbf{Pesticide residues} & \textbf{VMPR} \\
\midrule
Number of measurements    & 4,344,679     & 306,581,508    & 81,343,724 \\
Number of unique samples  & 904,607       & 8,789,645      & 5,482,221  \\
\bottomrule
\end{tabular}
\end{table}

\begin{table}[htbp]
\caption{Sample number distribution over sampling strategy. }
\label{tableA3}
\centering
\footnotesize
\begin{tabular}{p{3cm} R{3cm} R{2cm}}
\toprule
\textbf{Sampling strategy}        & \textbf{Number of samples} & \textbf{Percentage} \\
\midrule
objective sampling       & 8,059,735   & 53.10\% \\
selective sampling       & 3,553,278   & 23.40\% \\
other                    & 2,737,566   & 18.00\% \\
suspect sampling         & 811,145     & 5.30\% \\
convenient sampling      & 13,223      & 0.10\% \\
not specified            & 1,526       & 0.00\% \\
\textbf{TOTAL}           & \textbf{15,176,473} & \textbf{100.00\%} \\
\bottomrule
\end{tabular}
\end{table}

\newpage
\begin{table}[htbp]
\caption{Distribution of number of analytical results for chemical residues across different sampling strategies between 2015 - 2019.}
\label{tableA4}
\centering
\footnotesize
\begin{tabular}{p{3 cm} R{1cm} R{2.5 cm} R{2.5 cm} R{2.5 cm}}
\toprule
\textbf{Sampling strategy}     & 
\textbf{Year} & 
\textbf{Total number of analytical results} &
\textbf{Number of non-compliant results} & 
\textbf{Percentage of non-compliance} \\
\midrule
convenient sampling & 2015 & 0     & 0     & 0.0\% \\
                             & 2016 & 2,625  & 845   & 32.2\% \\
                             & 2017 & 4,235  & 1,322  & 31.2\% \\
                             & 2018 & 5,075  & 1,978 & 39.0\% \\
                             & 2019 & 15,810 & 1,947 & 12.3\% \\
objective sampling  & 2015 & 7,099  & 12    & 0.2\% \\
                             & 2016 & 7,132  & 0     & 0.0\% \\
                             & 2017 & 8,761  & 1     & 0.0\% \\
                             & 2018 & 11,991 & 17    & 0.1\% \\
                             & 2019 & 218,755 & 423  & 0.2\% \\
other               & 2015 & 2     & 0     & 0.0\% \\
                             & 2016 & 1,712  & 0     & 0.0\% \\
                             & 2017 & 24    & 0     & 0.0\% \\
                             & 2018 & 67    & 0     & 0.0\% \\
                             & 2019 & 4,983  & 24    & 0.5\% \\
selective sampling  & 2015 & 4,610  & 17    & 0.4\% \\
                             & 2016 & 6,748  & 1     & 0.0\% \\
                             & 2017 & 11,023 & 23    & 0.2\% \\
                             & 2018 & 11,953 & 50    & 0.4\% \\
                             & 2019 & 375,751 & 895  & 0.2\% \\
suspect sampling    & 2015 & 54    & 0     & 0.0\% \\
                             & 2016 & 68    & 0     & 0.0\% \\
                             & 2017 & 142   & 0     & 0.0\% \\
                             & 2018 & 329   & 0     & 0.0\% \\
                             & 2019 & 61,838 & 342   & 0.6\% \\
\bottomrule
\end{tabular}
\end{table}

\clearpage

\begin{table}[htbp]
\caption{Proportion of samples with unknown country origins over years.}
\label{tableA5}
\centering
\footnotesize
\begin{tabular}{p{1cm} R{2cm} R{2.5cm} R{2cm}}

\toprule

Year & Total number of samples & Number of samples with unknown   country origin & Unknown percentage \\
\midrule
1970 & 2         & 0       & 0.00\%  \\
1998 & 114       & 0       & 0.00\%  \\
1999 & 394       & 0       & 0.00\%  \\
2000 & 2,156     & 66      & 3.10\%  \\
2001 & 2,613     & 62      & 2.40\%  \\
2002 & 3,122     & 112     & 3.60\%  \\
2003 & 2,845     & 47      & 1.70\%  \\
2004 & 4,111     & 93      & 2.30\%  \\
2005 & 4,339     & 138     & 3.20\%  \\
2006 & 3,388     & 212     & 6.30\%  \\
2007 & 6,245     & 306     & 4.90\%  \\
2008 & 4,747     & 191     & 4.00\%  \\
2009 & 4,173     & 155     & 3.70\%  \\
2010 & 6,484     & 293     & 4.50\%  \\
2011 & 85,621    & 2,683   & 3.10\%  \\
2012 & 82,997    & 4,218   & 5.10\%  \\
2013 & 85,052    & 3,049   & 3.60\%  \\
2014 & 88,329    & 4,627   & 5.20\%  \\
2015 & 89,334    & 4,788   & 5.40\%  \\
2016 & 90,518    & 6,331   & 7.00\%  \\
2017 & 823,455   & 81,015  & 9.80\%  \\
2018 & 958,450   & 56,416  & 5.90\%  \\
2019 & 1,809,767 & 100,275 & 5.50\%  \\
2020 & 2,037,231 & 36,870  & 1.80\%  \\
2021 & 2,462,824 & 35,434  & 1.40\%  \\
2022 & 2,885,182 & 34,901  & 1.20\%  \\
2023 & 3,633,381 & 33,831  & 0.90\%  \\
2024 & 514       & 175     & 34.00\% \\
\bottomrule
\end{tabular}
\end{table}

\end{document}